\documentclass[a4paper,11pt,aps,prb,eqsecnum,amsmath,amssymb]{revtex4}
\usepackage{graphicx}
\usepackage{dcolumn}
\usepackage{bm}
\usepackage{epsfig}
\newcommand{\diag}{{\rm diag\,}}

\newcommand{\Str}{{\rm Str\,}}
\newcommand{\Sdet}{{\rm Sdet\,}}
\newcommand{\UOSp}{{\rm UOSp\,}}
\newcommand{\U}{{\rm U\,}}
\newcommand{\Herm}{{\rm Herm\,}}
\newcommand{\e}{\leavevmode\hbox{\small1\kern-3.8pt\normalsize1}}

\begin{document}
\newtheorem{definition}{Definition}[section]
\newtheorem{assumption}[definition]{Assumption}
\newtheorem{theorem}[definition]{Theorem}
\newtheorem{lemma}[definition]{Lemma}
\newtheorem{corollary}[definition]{Corollary}

\title{Integration of Grassmann variables over invariant functions on flat superspaces}
\author{Mario Kieburg}
\email{mario.kieburg@uni-due.de}
\author{Heiner Kohler}\author{Thomas Guhr}
\affiliation{University Duisburg-Essen, Lotharstrasse 1,
47048 Duisburg, Germany}

\date{15 January 2009}

\begin{abstract}
 We study integration over functions on superspaces. These functions are invariant under a transformation which maps the whole superspace onto the part of the superspace which only comprises purely commuting variables. We get a compact expression for the differential operator with respect to the commuting variables which results from Berezin integration over all Grassmann variables. Also, we derive Cauchy--like integral theorems for invariant functions on supervectors and symmetric supermatrices. This extends theorems partly derived by other authors. As a physical application, we calculate the generating function of the one--point correlation function in random matrix theory. Furthermore, we give another derivation of supermatrix Bessel functions for $\U(k_1/k_2)$.\\
\textbf{Published in:} Journal of Mathematical Physics \textbf{50}, 013528 (2009)\\
\text{[DOI: 10.1063/1.3049630]}
\end{abstract}

\maketitle

\section{Introduction}

In random matrix theory and in the theory of disordered systems, the supersymmetry method is an indispensable tool to study correlation functions and other quantities. Various integral theorems exist in superanalysis which have no counterpart in ordinary analysis. Parisi and Sourlas\cite{ParSou79} were the first to give such a theorem as a dimensional reduction
\begin{equation}\label{1.1}
 \int f(\eta^{*}\eta)d[\eta]=\left.\frac{1}{2\pi}\frac{df}{dz}(z)\right|_{z=0}  ,
\end{equation}
where the integration of the complex Grassmann variables $\eta,\eta^{*}$ is defined by
\begin{equation}\label{1.2}
 \int \eta^n d\eta=\int (\eta^{*})^n d\eta^{*}=\frac{\delta_{n1}}{\sqrt{2\pi}}\ \ \ {\rm and}\ \ \ d[\eta]=d\eta d\eta^{*}\ ,\ n\in\{0,1\}.
\end{equation}
They related this feature to an invariance of the integrand with respect to a super-rotation, which preserves the length of a supervector ($x^2+\eta^{*}\eta$). Here $x$ is an invertible commuting variable which may have nilpotent terms. Efetov\cite{Efe83} also obtained such a theorem for functions on the set of self adjoint complex (1/1)--supermatrices which are invariant under the supergroup U(1/1) and which have zero boundary condition at infinity. He discussed that these integral theorems are also true for an integration over superfunctions which are invariant under the action of more general groups and applied these theorems for his calculations.\cite{Efe97}

The equivalence of a Grassmann integration with the action of a differential operator is well known. If the superfunction is expanded in a Taylor series, any symmetry must be reflected in the coefficients of this series. Rothstein\cite{Rot87} found that a change of variables when integrating over a superspace leads to differential operators, which are incorporated into the invariant measure. These differential operators are exponential functions of vector fields. For an integration over a rotation--invariant superfunction on symmetric supermatrices, one can change from Cartesian integration variables to eigenvalue--angle coordinates. Then, the superfunction is independent of the angles.  Thus, only differential operators remain which stem from the transformation in the sense of Rothstein. However, as the vector fields of such a coordinate transformation are very difficult to calculate, we do not pursue this route here.

Another approach is due to Wegner\cite{Weg83} (worked out in Refs. \onlinecite{Con88} and \onlinecite{ConGro89}), who generalized Efetov's result to the case of an integration over functions on sets of supermatrices which are invariant under the action of a discrete subgroup of a classical Lie supergroup. In their studies, the integration of the Grassmann variables gives a differential operator with respect to the ordinary variables similar to Eq. \eqref{1.1}. The integration over these commuting variables is performed in Ref. \onlinecite{ConGro89} and leads to a Cauchy--like  integral theorem.

Our approach extends and generalizes the work of Wegner.\cite{Weg83} We employ rather general projection properties of functions $f(x,\eta)$ on superspaces to derive compact expressions for integrals over Grassmann variables. These expressions only involve derivatives with respect to the commuting variables $x$ of $f(x,0)$.

The article is organized as follows. In Sec. II, we map the integration over all Grassmann variables of a function $f(x,\eta)$ onto the action of a differential operator with respect to the commuting variables. This differential operator is uniquely defined by the invariance class of the function. The derivation of this differential operator is rather general and applies to a wide class of functions.

Our result leads to integral theorems for supervectors and for supermatrices, as explained in Sec. III and Sec. IV, respectively. We also extend results  obtained for the supergroup $\U(k/k)$ in Ref. \onlinecite{ConGro89} to the supergroups $\U(k_1/k_2)$ and $\UOSp(k_1/k_2)$, respectively. In Sec. V, we apply our method on two examples. First we work out the generating function for the one--point correlation function\cite{Guh91,GuhKoh02b} in random matrix theory for the cases of Gaussian real orthogonally and unitarily invariant ensembles. Thereby, we rederive the supermatrix Bessel function for $\U(1/1)$(Refs. \onlinecite{Guh91},\onlinecite{Guh93}, and \onlinecite{Guh96}) and for $\UOSp(2/2)$.\cite{GuhKoh02b} As a second example we present a new derivation of the  supersymmetric Itzykson--Zuber integral with respect to $\U(k_1/k_2)$.\cite{Guh91,Guh96} There, we omit the boundary terms referred to the literature\cite{Guh91,Guh93,Guh06,BasAke07} as Efetov--Wegner terms. In Appendix A, we show that for superfunctions invariant under $\UOSp(1/2)$ no Cauchy--like integral theorem exists.

\section{Integration of Grassmann variables over invariant functions}

Following Berezin,\cite{Ber87} we consider a complex Grassmann algebra $\Lambda=\bigoplus\limits_{j=0}^{2L} \Lambda_j$. We have $L\in\mathbb{N}$ pairs of complex Grassmann variables $\{\eta_j,\eta_j^{*}\}_{1\leq j\leq L}$. We define in a canonical way the space of even $\Lambda^0=\bigoplus\limits_{j=0}^{L} \Lambda_{2j}$ and odd $\Lambda^1=\bigoplus\limits_{j=0}^{L-1} \Lambda_{2j+1}$ variables. Let $M$ be a real $p$-dimensional differentiable manifold. We are interested in functions on the superspace $\Lambda(p,2L)$ with the base $M$ and a sheaf of algebras $\mathfrak{A}$. Let $U$ be an open subset of $M$ then $\mathfrak{A}(U)$ is the algebra of functions on $U$ with values in $\Lambda$. We split $\Lambda(p,2L)$ into a direct sum $\Lambda^0(p,2L)\oplus\Lambda^1(p,2L)$ corresponding to the ${\mathbb Z}_2$ grading of $\Lambda$. Functions on $L(p,2L)=(\Lambda^0(p,2L))^p\times(\Lambda^1(p,2L))^{2L}$ with values in $\Lambda$ can be represented as a finite power series in the generators of $\Lambda$
\begin{equation}\label{2.1}
 f(x,\eta)=\sum\limits_{j_1,j_2\in \mathbb{I}}f_{j_1j_2}(x)\left(\prod\limits_{n=1}^L(\eta_n^{*})^{j_{1n}}\eta_n^{j_{2n}}\right),
\end{equation}
where $j_1$ and $j_2$ are multiple indices in the set $\mathbb{I}=\{0,1\}^L$ and $\prod\limits_{n=1}^L A_n=A_1A_2\hdots A_n$ is an ordered product. We call functions, which have the representation (\ref{2.1}), superfunctions.

Let $L(p,2L)$ be a Riemannian superspace with metric $g$. This metric is assumed to be diagonal and constant. The inner product of two elements in $L(p,2L)$ is
\begin{equation}\label{2.6}
 g((a,\alpha),(b,\beta))=\sum\limits_{n=1}^p g_n a_n b_n + \sum\limits_{m=1}^L h_m(\alpha_m^*\beta_m+\beta_m^*\alpha_m)\ ,\ g_n\in\mathbb{R}^+\ {\rm and}\ h_m\in\mathbb{C}.
\end{equation}
Lower case Latin letters such as $a,b,\hdots$ denote elements in $\Lambda^0$ and lower case Greek letters such as $\alpha,\beta,\hdots$ denote odd elements in $\Lambda^1$. Lower case Latin letters such as $x,y$ denote real numbers. 

The idea of Wegner\cite{Weg83} (worked out in Ref. \onlinecite{ConGro89}) is the following. Consider the superfunction $f$ which is invariant under a discrete subgroup of a Lie group acting on $L(p,2L)$, or equivalently, which is invariant under all transformations connecting the commuting with the anticommuting variables. Then all $f_{j_1j_2}$ are functionally dependend on the body $f_{0,0}(x)=f(x,0)=f(x)$. In this spirit, we assume that there exists a differentiable map $\phi:\ L(p,2L)\rightarrow L(p,2L)$ with the properties $\phi(a,\alpha)=(r(a,\alpha),0)$ and $f(x,\eta)=f(r(x,\eta),0)=f(r(x,\eta))$ ,where $\{r_j\}_{1\leq j\leq p}$ are mappings onto $\Lambda^0$. The image $\mathcal{N}=\{(r(x,\eta),0)|(x,\eta)\in L(p,2L)\}\subset(\Lambda^0(p,2L))^p$ is a differentiable subsupermanifold of $(\Lambda^0(p,2L))^p$. The dimension of this image tells us how many independent variables are needed to describe the resulting supermanifold in terms of commuting variables. These variables are referred to as radial variables. The remaining variables which parametrize the complement of this submanifold with respect to  $L(p,2L)$ are referred to as angular variables.
\begin{theorem}\label{t1}\ \\
 Let $f$ be a differentiable superfunction on $L(p,2L)$ which is invariant under a differentiable map $\phi:\ L(p,2L)\rightarrow L(p,2L)$ with $\phi(a,\alpha)=(r(a,\alpha),0)$, that is $f(x,\eta)=f(r(x,\eta))$. Define the integral of $f$ with respect to 
all Grassmann variables contained in $f$ by $\int f(x,\eta) d[\eta]$, with the measure $d[\eta]=\prod\limits_{n=1}^Ld[\eta_n]$. 
Then there exists a differential operator $D_{C,S}$ with respect to the real variables such that
\begin{equation}\label{t1.1}
 \int f(x,\eta)d[\eta]=D_{C,S}(r)f(r,0)
\end{equation}
holds. The differential operator $D_{C,S}$ is explicitly given by  
\begin{equation}\label{t1.2}
 D_{C,S}(r)=\frac{1}{L!(4\pi)^L}\left(\prod\limits_{m=1}^L h_m\right)\sum\limits_{n=0}^L\binom{L}{n}\Delta_C^{L-n}\left(-\Delta_{S,r}(r)\right)^n,
\end{equation}
where
\begin{equation}\label{2.8}
 \Delta_C=\sum\limits_{n=1}^p\frac{1}{g_n}\frac{\partial^2}{\partial x_n^2}\ \ \ {\rm and}\ \ \ \Delta_S=\Delta_C+2\sum\limits_{m=1}^L\frac{1}{h_m}\frac{\partial^2}{\partial \eta_m\partial\eta_m^*}
\end{equation}
are the Laplace--Beltrami operators of the pure commuting part and the whole superspace, respectively. Thus, the index $C$ refers to a differential operator acting only on the space $(\Lambda^0(p,2L))^p$ and the index $S$ denotes differential operators acting on the whole space $L(p,2L)$. $\Delta_{S,r}(r)$ is the radial part of the differential operator $\Delta_S$ and is described in the proof below.
\end{theorem}
\textbf{Proof:}\\
One can project $f$ onto all $f_{j_1j_2}$ with a projector $P_{j_1j_2}$ in the following way. The operator identity
\begin{equation}\label{2.2}
 1=\frac{\partial}{\partial\eta_n}\eta_n+\eta_n\frac{\partial}{\partial\eta_n}
\end{equation}
holds for every Grassmann generator $\eta_n$ and also for its complex conjugate $\eta_n^*$. The first term on the right hand side projects onto the function $f(x,\eta_1,\hdots,\eta_{n-1},0,\eta_{n+1},\hdots)$ and the second term projects, up to a sign, onto $\eta_n (df/dz)(x,\eta_1,\hdots,\eta_{n-1},z,\eta_{n+1},\hdots)|_{z=0}$. Thus, the generalization of Eq. \eqref{2.2} to all generators is
\begin{equation}\label{2.3}
 \begin{split}
  1  = & \prod\limits_{n=1}^L\left(\frac{\partial}{\partial\eta_n}\eta_n+\eta_n\frac{\partial}{\partial\eta_n}\right)\left(\frac{\partial}{\partial\eta_n^*}\eta_n^*+\eta_n^*\frac{\partial}{\partial\eta_n^*}\right)=\\
     = & \sum\limits_{j_1,j_2\in \mathbb{I}} (-1)^{J(j_1,j_2)}\left(\prod\limits_{n=1}^L(\eta_n^*)^{j_{1n}}\eta_n^{j_{2n}}\right)\left(\prod\limits_{n=1}^L\frac{\partial^2}{\partial\eta_n\partial\eta_n^*}\right)\left(\prod\limits_{n=1}^L(\eta_n^*)^{1-j_{1n}}\eta_n^{1-j_{2n}}\right) ,
 \end{split}
\end{equation}
where
\begin{equation}
 J(j_1,j_2)=\sum\limits_{n=1}^L(1-j_{1n})j_{2n}+\sum\limits_{1\leq n< m\leq L}(j_{1n}+j_{2n})(j_{1m}+j_{2m}).
\end{equation}
The action of this operator onto the left hand side of \eqref{2.1} gives an expression for $f$ which can be compared term by term with the right hand side of \eqref{2.1}. This yields
\begin{equation}\label{2.4}
 f_{j_1j_2}(x)=(-1)^{J(j_1,j_2)}\left(\prod\limits_{n=1}^L\frac{\partial^2}{\partial\eta_n\partial\eta_n^*}\right)\left(\prod\limits_{n=1}^L(\eta_n^*)^{1-j_{1n}}\eta_n^{1-j_{2n}}\right)f(x,\eta).
\end{equation}
We define the projector onto the body $f_{0,0}$ of such a superfunction,
\begin{equation}\label{2.5}
 P_{0,0}=\left(\prod\limits_{n=1}^L \frac{\partial^2}{\partial\eta_n\partial\eta_n^*}\right)\left(\prod\limits_{n=1}^L \eta_n^*\eta_n\right).
\end{equation}
An application of $P_{0,0}$ on $f_{j_1j_2}(x)$ is the identity because $f_{j_1j_2}$ does not depend on any Grassmann variable. However, the action of $P_{1,\hdots,1}$ onto $f_{j_1j_2}$ is zero. When we analyze the projector $P_{1,\hdots,1}$, i.~e.~the first part on the left hand side of Eq. \eqref{2.5}, we must take into account the commutating variables. We obtain
\begin{eqnarray}
  P_{1,\hdots,1}&=&\prod\limits_{n=1}^L \frac{\partial^2}{\partial\eta_n\partial\eta_n^*}=  \frac{(-2)^{-L}}{L!}\prod\limits_{m=1}^L h_m\left(2\sum\limits_{n=1}^L\frac{1}{h_n}\frac{\partial^2}{\partial\eta_n^*\partial\eta_n}\right)^L= \frac{2^{-L}}{L!}\prod\limits_{m=1}^L h_m\left(\Delta_C-\Delta_S\right)^L=\nonumber\\
  &= & \frac{2^{-L}}{L!}\prod\limits_{m=1}^L h_m\sum\limits_{n=0}^L\binom{L}{n}\Delta_C^{L-n}\left(-\Delta_S\right)^n .\label{2.7}
\end{eqnarray}
 We plug Eq. \eqref{2.7} into \eqref{2.4} and act with the operator $P_{0,0}$ from the left. We use that $P_{0,0}$ commutes with $\Delta_C$ and obtain
\begin{equation}\label{2.9}
 f_{j_1j_2}(x)=\frac{(-1)^{J(j_1,j_2)}2^{-L}}{L!}\prod\limits_{m=1}^L h_m\sum\limits_{n=0}^L\binom{L}{n}\Delta_C^{L-n}P_{0,0}\left(-\Delta_S\right)^n\left(\prod\limits_{n=1}^L(\eta_n^*)^{1-j_{1n}}\eta_n^{1-j_{2n}}\right)f(x,\eta).
\end{equation}

We now choose the metric $g$ such that the Laplacian $\Delta_S$ can be written in radial and angular coordinates. Then it splits into a sum of two differential operators $\Delta_{S,r}+\Delta_{S,\varphi}$. The radial part $\Delta_{S,r}$ contains only radial coordinates and partial derivatives thereof. The angular part $\Delta_{S,\varphi}$ only contains partial derivatives with respect to the angular coordinates. The action of $\Delta_{S,\varphi}$ onto an invariant function is zero. The radial part fulfills
\begin{equation}\label{2.10}
 P_{0,0}\Delta_{S,r}(r(x,\eta))=\Delta_{S,r}(r(x,0))P_{0,0}=\Delta_{S,r}(r(x))P_{0,0}.
\end{equation}
We summarize these results and apply them onto the integral over an invariant function,
\begin{equation}\label{2.11}
 \begin{split}
  \int f(x,\eta)d[\eta]= & \frac{1}{(2\pi)^L}f_{1,\hdots,1}(x)=\\
  \overset{\eqref{2.9}}{=} & \frac{1}{L!(4\pi)^L}\prod\limits_{m=1}^L h_m\sum\limits_{n=0}^L\binom{L}{n}\Delta_C^{L-n}P_{0,0}\left[-\Delta_{S}(r(x,\eta))\right]^nf(x,\eta)=\\
  = & \frac{1}{L!(4\pi)^L}\prod\limits_{m=1}^L h_m\sum\limits_{n=0}^L\binom{L}{n}\Delta_C^{L-n}P_{0,0}\left[-\Delta_{S,r}(r(x,\eta))\right]^nf(r(x,\eta))=\\
  \overset{\eqref{2.10}}{=} & \frac{1}{L!(4\pi)^L}\prod\limits_{m=1}^L h_m\sum\limits_{n=0}^L\binom{L}{n}\Delta_C^{L-n}\left[-\Delta_{S,r}(r(x))\right]^nf(r(x))=\\
  = & \left.D_{C,S}(r)f(r)\right|_{r=r(x)}.\hspace*{7cm}\square
 \end{split}
\end{equation}

This formula will be applied to special cases in the ensuing sections. We will thereby also rederive some results of Refs. \onlinecite{Weg83,Con88,ConGro89}. However, in particular in the matrix case, to be discussed in Sec.~\ref{matrices}, the operator $D_{C,S}(r)$ becomes very complex and a handier expression is thus highly desirable. We therefore first rewrite it, using a transformation akin to the Baker--Campbell--Hausdorff formula.
\begin{lemma}\label{l1}\ \\
 The operator $D_{C,S}(r)$ is a differential operator of order $L$ and can be written as
 \begin{equation}\label{l2.1}
 D_{C,S}(r)=\frac{1}{L!(4\pi)^L}\left(\prod\limits_{m=1}^L h_m\right){\rm IAd}[\Delta_C,\Delta_C-\Delta_{S,r}]^L(\textbf{1})
 \end{equation}
 where
 \begin{eqnarray}\label{l2.2}
  {\rm IAd}[A,B](H)&=&[A,H]_-+HB\ \ \ {\rm and}\nonumber\\
 {\rm IAd}[A,B]^N(H)&=&[A,{\rm IAd}[A,B]^{N-1}(H)]_-+{\rm IAd}[A,B]^{N-1}(H)B
 \end{eqnarray}
 for three arbitrary linear operators $A,B$ and $H$ and $\textbf{1}$ is the identity operator. The operator $D_{C,S}(r)$ is 
a differential operator of order $L$.
\end{lemma}
We use the symbol ${\rm IAd}$ to indicate a similarity to the linear operator ${\rm Ad}[A](B)=[A,B]_-=AB-BA$ which is the adjoint representation of a linear operator $A$.\\
\textbf{Proof:}\\
Consider two noncommuting finite dimensional real matrices $A$ and $B$, then
\begin{equation}\label{2.12}
  \sum\limits_{n=0}^L\binom{L}{n}A^{L-n}(B-A)^n= \left.\frac{d^L}{dt^L}\left(e^{At}e^{(B-A)t}\right)\right|_{t=0}.
\end{equation}
Using $\phi(s)=e^{Ats}e^{(B-A)ts}$ we obtain
\begin{equation}\label{2.13}
  \frac{d}{ds}\phi(s)=\left([A,\phi(s)]_-+\phi(s)B\right)t\ \quad  {\rm and}\ \ \phi(0)=1.
\end{equation}
Consequently, we find
\begin{equation}\label{2.14}
  \phi(s)=e^{st\ {\rm IAd}[A,B]}(\textbf{1})
\end{equation}
and arrive at
\begin{equation}\label{2.15}
  \sum\limits_{n=0}^L\binom{L}{n}A^{L-n}(B-A)^n= {\rm IAd}[A,B]^L(\textbf{1}).
\end{equation}
Since this formula is a finite polynomial in the operators $A$ and $B$, Eq. \eqref{2.15} holds for any linear operator. This means that we can perform this rearrangement for the Laplacians in the operator $D_{C,S}(r)$.

The operator $(\Delta_C-\Delta_{S,r})=P_{0,0}(\Delta_C-\Delta_S+\Delta_{S,\phi})$ is a differential operator of order $1$ because the Grassmann variables can be viewed as a perturbative term in the pure commutative part $(\Lambda^0(p,2L))^p$. The flat operator $(\Delta_S-\Delta_C)$ contains only second derivatives with respect to the Grassmann variables. The derivative of a Grassmann variable reads in radial--angle coordinates
\begin{equation}\label{h2.0}
 \frac{\partial}{\partial \eta_n}=\sum\limits_{m=1}^{{\rm dim}(\mathcal{N})}\frac{\partial r_m}{\partial \eta_n}\frac{\partial}{\partial r_m}+\sum\limits_{m=1}^{p+2L-{\rm dim}(\mathcal{N})}\frac{\partial \phi_m}{\partial \eta_n}\frac{\partial}{\partial \phi_m}.
\end{equation}
$\partial r_m/\partial \eta_n$ is anticommuting and satisfies $P_{0,0}\partial r_m/\partial \eta_n=0$. Thus, $D_{C,S}(r)$ is a differential operator of order $L$.\hfill$\square$\\

\section{Integral theorems for invariant functions on supervectors}

As a first example for Theorem \ref{t1}, we consider functions on a space of supervectors which are invariant under the action of ${\rm UOSp}^{(+)}(p/2L)$. The index ``$+$'' refers to the classification of Riemannian symmetric superspaces by Zirnbauer\cite{Zir96} and corresponds to the notation of Ref. \onlinecite{KohGuh05}. It indicates that the boson--boson block has real entries and the fermion--fermion block has quaternionic entries.  ${\rm UOSp}^{(+)}(p/2L)$ is a representation of ${\rm UOSp}(p/2L)$. An alternative notation is used by Zirnbauer\cite{Zir96}. In our notation ${\rm UOSp}^{(-)}(2p/2L)$ denotes the case in which the boson--boson block is quaternionic and the fermion--fermion block is real.

The invariance condition on a supervector $v=(a_1,\hdots,a_p,\alpha_1,\hdots,\alpha_L,\alpha_1^*,\hdots,\alpha_L^*)^T$ is
\begin{equation}\label{3.1}
 f(v)=f(Uv)
\end{equation}
for all $U\in \UOSp^{(+)}(p/2L)$. In this case the metric (\ref{2.6}) is given by $g_n=h_m=1$. The function $f$ only depends on the invariant length $r=\sqrt{v^\dagger v}$. Equation \eqref{t1.1} takes the form
\begin{equation}\label{3.2}
   \int f(x,\eta)d[\eta]=
  \frac{1}{L!(4\pi)^L}\sum\limits_{n=0}^L\binom{L}{n}\left(\frac{1}{r^{p-1}}\frac{\partial}{\partial r}r^{p-1}\frac{\partial}{\partial r}\right)^{L-n}\left(-\frac{1}{r^{p-1-2L}}\frac{\partial}{\partial r}r^{p-1-2L}\frac{\partial}{\partial r}\right)^nf(r).
\end{equation}

The differential operator on the right hand side of \eqref{3.2} is independent of $p$ due to the invariance of $f$ with respect to the orthogonal group ${\rm O}(p)$ in the commuting part. Hence, the commuting variables on the left hand side of \eqref{3.2} can be written in terms of the radial coordinates only. We calculate the integral over $f$ on an effective superspace $\Lambda^0(1,L)$ and obtain
\begin{equation}\label{3.3}
  \int f(x,\eta)d[\eta]=\frac{1}{L!(4\pi)^L}\sum\limits_{n=0}^L\binom{L}{n}
                           \left(\frac{\partial^2}{\partial r^2}\right)^{L-n}\left(-r^{2L}
                            \frac{\partial}{\partial r}\frac{1}{r^{2L}}\frac{\partial}{\partial r}\right)^nf(r)=D_{r}^{(1,L)}f(r) .
\end{equation}
In particular, we obtain for small dimension $L$
\begin{eqnarray}
 & & L=1:\label{3.4}
            	 \int f(x,\eta)d[\eta]=\frac{1}{2\pi}\frac{1}{r}\frac{\partial}{\partial r}f(r),\\
 & & L=2:\label{3.5}
            	 \int f(x,\eta)d[\eta]=\left(\frac{1}{2\pi }\right)^2\left(\frac{1}{r^2}\frac{\partial^2}{\partial r^2}-\frac{1}{r^3}\frac{\partial}{\partial r}\right)f(r).
\end{eqnarray}
In general, we can formulate the following integral theorem. This theorem generalizes the theorem of Wegner\cite{Weg83} worked out in Theorem 4.1 of Ref. \onlinecite{ConGro89}, which focuses on complex supervectors, to the case of real supervectors.
\begin{theorem}[Real supervectors]\label{t2}\ \\
 Let $f$ be a differentiable function of supervectors $v=(a_1,\hdots,a_p,\alpha_1,\hdots,\alpha_L,\alpha_1^*,\hdots,\alpha_L^*)^T$ 
and of its adjoint $v^{\dagger}$. Let $f$ be invariant under the action of $\UOSp^{(+)}(p/2L)$ and have zero boundary condition at infinity. Then, we have
 \begin{equation}\label{3.6}
  \int\limits_{\mathbb{R}^{p}} \int f(x,\eta)d[\eta]d[x]=
  \begin{cases}
   \displaystyle\imath^p\left.\left(\frac{1}{\pi r}\frac{\partial}{\partial r}\right)^{L-p/2}f(r)\right|_{r=0} & ,\ p<2L\ \wedge\ p\in\left(2\mathbb{N}_0\right)\\
   \displaystyle\frac{\imath^{p-1}}{\pi}\int\limits_{\mathbb{R}}\left(\frac{1}{\pi \tilde{x}}\frac{\partial}{\partial \tilde{x}}\right)^{L-(p-1)/2}f(\tilde{x})d\tilde{x} & ,\ p<2L\ \wedge\ p\in\left(2\mathbb{N}_0+1\right)\\
   \displaystyle\left(-1\right)^{L}f(0) & ,\ p=2L\\
   \displaystyle\left(-1\right)^{L}\int\limits_{\mathbb{R}^{p-2L}}f(\tilde{x})d[\tilde{x}] & ,\ p>2L
  \end{cases}
 \end{equation}
 where $d[x]=\prod\limits_{n=1}^{p}dx_n$ and $\tilde{x}$ refers to the canonical embedding of the lower dimensional integration set in $L(p,2L)$.
\end{theorem}
\textbf{Proof:}\\
The differential operator in Eq.~\eqref{3.3} can be written as
\begin{equation}\label{p3.1}
 D_r^{(1,L)}=\left(\frac{1}{2\pi}\right)^{L}\left(\frac{1}{r}\frac{\partial}{\partial r}\right)^L.
\end{equation}
This follows from the commutation relation
\begin{equation}\label{p3.0}
 \left[\frac{\partial^2}{\partial r^2},\frac{1}{r}\frac{\partial}{\partial r}\right]_-=-2\left(\frac{1}{r}\frac{\partial}{\partial r}\right)^2
\end{equation}
and from Eq. \eqref{l2.1}. One can also expand $f\left(\sqrt{r^2+2\sum\limits_{n=1}^L\eta_n^*\eta_n}\right)$ in a Taylor expansion in $r^2$ and take into account only the highest term of the power series in the Grassmann variables. For $p=2L$ one finds
\begin{equation}\label{p.3.2}
 \begin{split}
  \int\limits_{\mathbb{R}^{2L}} \int f(x,\eta)d[\eta]d[x]= & \int\limits_{\mathbb{R}^{2L}}\left(\frac{1}{2\pi}\right)^{L}\left(\frac{1}{r}\frac{\partial}{\partial r}\right)^L f(x)d[x]=\\
  = & \frac{2}{2^L(L-1)!}\int\limits_{\mathbb{R}^{+}}r^{2L-1}\left(\frac{1}{r}\frac{\partial}{\partial r}\right)^L f(r)dr=\\
  = & \frac{1}{(L-1)!}\int\limits_{\mathbb{R}^{+}}\left(r^2\right)^{L-1}\left(\frac{\partial}{\partial (r^2)}\right)^L f\left(\sqrt{r^2}\right)dr^2=\\
  = &\left(-1\right)^{L}f(0).
 \end{split} 
\end{equation}
This proves the third equation in Eq.~\eqref{3.6}. To prove the first equation in Eq.~\eqref{3.6}, we expand the superfunction $f$ in a power series in $p$ pairs of Grassmann variables. We integrate over all real variables and over these pairs and apply the third equation of Eq.~\eqref{3.6}. We can then expand the rest of the Grassmann variables since we know that $f$ only depends on the length of the remaining supervector. This proves the first equation in Eq.~\eqref{3.6}. 

The second equation in Eq.~\eqref{3.6} can be treated similarly, however, now there is an additional real variable which can not be integrated. We consider $p=1$ and $L=2$. Then, we have to integrate
\begin{equation}\label{h4.0}
 \int\limits_{\mathbb{R}}\frac{1}{r}\frac{\partial}{\partial r}f(r)dr.
\end{equation}
We need an $r$ in the numerator for cancellation with the singular term $r^{-1}$. Such a contribution is guaranteed if we have to every pair of Grassmann variables a pair of real variables.

The same reasoning applies for the fourth equation in Eq.~\eqref{3.6}. Here we do not find for every pair of real variables a pair of Grassmann variables. Thus, we can integrate over $2L$ real variables using the third equation of Eq.~\eqref{3.6} and we are left with an integral over $(p-2L)$ real variables.\hfill$\square$

Theorem \ref{t2} can easily be extended to functions $f(v,v^{\dagger})$ of a complex supervector $v=(a_1+\imath b_1,\hdots,a_p+\imath b_p,\alpha_1,\hdots,\alpha_L)^T$ and of its adjoint $v^\dagger$, which are invariant under the action of the group $\U(p/L)$. The radial variable is the length $\sum\limits_{n=1}^p(a^2+b^2)+\sum\limits_{m=1}^L\alpha_m^*\alpha_m$ of the supervector. The metric $g$, Eq. \eqref{2.6}, is defined by $g_n=1$ and $h_m=1/2$. We find
\begin{equation}\label{3.7}
 \int f(x,y,\eta)d[\eta]= \left(\frac{1}{4\pi}\right)^{L}\left(\frac{1}{r}\frac{\partial}{\partial r}\right)^Lf(r)=D_{r}^{(2,L)}f(r).
\end{equation}
An integration theorem follows right away.
\begin{theorem}[Complex supervectors]\label{t3}\ \\
 Let $f$ be a differentiable function on supervectors $v=(a_1+\imath b_1,\hdots,a_p+\imath b_p,\alpha_1,\hdots,\alpha_L)^T$ and of its adjoint $v^\dagger$, which is invariant under the action of $\U(p/L)$ and which has zero boundary condition at infinity, then 
 \begin{equation}\label{3.8}
  \int\limits_{\mathbb{C}^{p}} \int f(z,\eta)d[\eta]d[z]=
  \begin{cases}
   \displaystyle(-1)^p\left.\left(-\frac{1}{2}\right)^{L}\left(\frac{1}{\pi r}\frac{\partial}{\partial r}\right)^{L-p}f(r)\right|_{r=0} & ,\ p<L\\
   \displaystyle\left(-\frac{1}{2}\right)^{L}f(0) & ,\ p=L\\
   \displaystyle\left(-\frac{1}{2}\right)^{L}\int\limits_{\mathbb{C}^{p-L}}f(\tilde{z})d[\tilde{z}] & ,\ p>L\\
  \end{cases}
 \end{equation}
 where $d[z]=\prod\limits_{n=1}^{p}d{\rm Re}\ z_nd{\rm Im}\ z_n$ and $\tilde{z}$ refers to the canonical embedding of the lower dimensional integration set in $L(2p,2L)$.
\end{theorem}
For $p=L$, this integral theorem coinides with the integral Theorem 4.1  of  Ref. \onlinecite{ConGro89}.

We now turn to functions $f(v,v^{\dagger})$ of vectors $v=(A_1, \hdots,A_p,\mathfrak{A}_1,\hdots,\mathfrak{A}_L )^T$ over the field of quaternions $A_n=\begin{bmatrix} a_n+\imath b_n & c_n+\imath d_n \\ -c_n+\imath d_n & a_n-\imath b_n \end{bmatrix}$ and $\mathfrak{A}_m=\begin{bmatrix} \alpha_m & \alpha_m^* \\ \beta_m & \beta_m^* \end{bmatrix}$. We assume these function to be invariant under $\UOSp^{(-)}(2p/2L)$. Hence, these functions only depend on the quaternionic matrix $v^{\dagger}v$ which is diagonal and self-adjoint. The corresponding metric is defined by $g_n=2$ and $h_m=1$. This leads to the differential operator
\begin{equation}\label{3.9}
 \int f(x,y,\eta)d[\eta]=\left(\frac{1}{4\pi}\right)^{L}\left(\frac{1}{r}\frac{\partial}{\partial r}\right)^Lf(r)=D_{r}^{(4,L)}f(r),
\end{equation}
implying the integral theorem
\begin{theorem}[Quaternionic supervectors]\label{t4}\ \\
 Let $f$ be a differentiable function on $(2p/2L)\times2$--supervectors of the form $v=(A_1, \hdots,A_p,\mathfrak{A}_1,\hdots,\mathfrak{A}_L )^T$ with $A_n$ quaternionic and $\mathfrak{A}_m=\begin{bmatrix} \alpha_m & \alpha_m^* \\ \beta_m & \beta_m^* \end{bmatrix}$ and its adjoint $v^\dagger$ which is invariant under the action of $\UOSp^{(-)}(2p/2L)$ and has zero boundary condition at infinity. Then, we have
 \begin{equation}\label{3.10}
 \int\limits_{\mathbb{H}^{p}} \int f(A,\eta)d[\eta]d[A]=
 \begin{cases}
   \displaystyle\frac{1}{4^L}\left.\left(\frac{1}{\pi r}\frac{\partial}{\partial r}\right)^{2(L-p)}f(r)\right|_{r=0} & ,\ p<L \\
   \displaystyle\frac{1}{4^L}f(0) & ,\ p=L \\
   \displaystyle\frac{1}{4^L}\int\limits_{\mathbb{H}^{p-L}}f(\widetilde{A})d[\widetilde{A}] & ,\ p>L
  \end{cases}
 \end{equation}
 where $d[A]=\prod\limits_{n=1}^{p}d{\rm Re}\ z_{1n}d{\rm Im}\ z_{1n}d{\rm Re}\ z_{2n}d{\rm Im}\ z_{2n}$ and $A_n=\begin{bmatrix} z_{n1} & z_{n2} \\ -z_{n2}^* & z_{n1}^* \end{bmatrix}$ and $\widetilde{A}$ refers to the canonical embedding of the lower dimensional integration set in $L(4p,4L)$.
\end{theorem}
The three theorems given here are crucial for the proof of the ensuing integral theorems for invariant functions on supermatrix spaces.

\section{Integral theorems for invariant functions on supermatrices}
\label{matrices}

To begin with, we consider functions of $(k_1/k_2)$--supermatrices invariant under the action of $\U(k_1/k_2)$. In this case the integral theorem is equivalent to the one obtained in Refs. \onlinecite{Weg83} and \onlinecite{ConGro89}. However, there the authors did not derive the differential operator we will present here. Furthermore, we will study $\U(k_1/k_2)$--symmetric supermatrices with a generalized Wick rotation by a real phase $\psi\in]0,\pi[$. A $\U(k_1/k_2)$--symmetric supermatrix $\Sigma$ is form invariant under the action of $\U(k_1/k_2)$ and reads in boson--fermion block notation
\begin{equation}\label{4.1}
 \Sigma=\begin{bmatrix} \Sigma_1 & e^{\imath\psi/2}\alpha^{\dagger} \\ e^{\imath\psi/2}\alpha & e^{\imath\psi}\Sigma_2 \end{bmatrix}
\end{equation}
where the body of $\Sigma_1$ is a Hermitian $k_1\times k_1$--matrix, the body of $\Sigma_2$ is a Hermitian $k_2\times k_2$--matrix and $\alpha$ is a $k_1\times k_2$--matrix with anticommuting entries with no further symmetry. Lower case Greek letters such as $\sigma,\rho,\hdots$ denote supermatrices where the entries of boson--boson and fermion--fermion block are elements in $\Lambda_0$ and the entries of the boson--fermion block are elements in $\Lambda_1$. If the entries of the supermatrix blocks generally lie in $\Lambda^0$ or $\Lambda^1$, respectively, then these supermatrices are denoted with upper case Greek letters such as $\Sigma,\Xi,\hdots$\ .

We introduce a generalized Wick rotation for two reasons. First, we will need a complex phase for the Wick rotation, which differs from the choice $\pm 1$ in the derivation of integral theorems. Second, in random matrix theory such phases guarantee convergence of integrals over superfunctions which are extensions of characteristic functions in superspace. The choice $e^{\imath\psi} =\imath$ has been employed of instance in Refs. \onlinecite{Efe83} and \onlinecite{Guh06}. However, other choices are important as well. For example, the integral over the superfunction $e^{-\Str{\Sigma^4}}$ is not convergent if we use the Wick rotation $\psi = \pi/2$.

Let $\Sigma$ and $\Xi$ be two $\U(k_1/k_2)$--symmetric supermatrices as given in Eq.~(\ref{4.1}), then the metric is defined through the supertrace\cite{Ber87} $g(\Sigma,\Xi)=\Str\Sigma\Xi$. We notice that the body of this metric does not lie in $\mathbb{R}^+$ for an arbitrary Wick rotation. However, the body of $g(\Sigma,\Xi)$ lies for two $\U(k_1/k_2)$--symmetric supermatrices with Wick rotation $\imath$ in $\mathbb{R}^+$. The metric is for such a choice
\begin{equation}\label{h4.1}
 g_n=\begin{cases}
      	1	&	,\ n\text{ is a diagonal index,}\\
	2	&	,\ n\text{ is an off-diagonal index}
     \end{cases}\ \ {\rm and}\ \ h_m=\imath.
\end{equation}
We continue this metric in an analytic way on the space of $\U(k_1/k_2)$--symmetric supermatrices with arbitrary Wick rotation. We exchange the real numbers of the fermion--fermion block entries to $-\imath e^{\imath\psi}$ times the real numbers and substitute the Grassmann variables with $\sqrt{-\imath e^{\imath\psi}}$ times the same Grassmann variables.

Now let $f$ be an invariant function on the space of $\U(k_1/k_2)$--symmetric supermatrices,
\begin{equation}
\label{4.1a}
f(\Sigma) = f(U^{-1}\Sigma U) \ ,\ \quad U \in  \U(k_1/k_2) .
\end{equation}
We identify the radial part of the space of $\U(k_1/k_2)$--symmetric supermatrices as the space of diagonal matrices $s=\diag(s_{1,1},\hdots,s_{k_1,1},e^{\imath\psi}s_{1,2},\hdots,e^{\imath\psi}s_{k_2,2})$. Therefore, we can apply Theorem \ref{t1} and find
\begin{eqnarray}\label{4.2}
\int f(\sigma)d[\eta] & 
  = & \frac{e^{\imath\psi k_1k_2}}{(k_1k_2)!(4\pi)^{k_1k_2}}\sum\limits_{n=0}^{k_1k_2}\binom{k_1k_2}{n}
         \left(\Delta_{s_1}^{(2;k_1)}-e^{-2\imath\psi}\Delta_{s_2}^{(2;k_2)}\right)^{k_1k_2-n}
         \left(-\Delta_{s}^{(2,2;k_1k_2)}\right)^nf(s)= \nonumber\\
                       &=& D_{s}^{(2,2;k_1k_2)}f(s)
\end{eqnarray}
where
\begin{equation}\label{4.3}
 \Delta_{s}^{(2;k)}=\sum\limits_{j=1}^{k}\frac{1}{\Delta_k^2(s)}\frac{\partial}{\partial s_j}\Delta_k^2(s)\frac{\partial}{\partial s_j}
\end{equation}
is the radial part of the Laplace operator on the space of ordinary Hermitian matrices. Here, $\Delta_k(s)=\det[s_n^{m-1}]_{1\leq n,m\leq k}$ is the Vandermonde determinant. We denote by $\Delta_{s}^{(2,2;k_1k_2)}$ the radial part of the Laplacian in the superspace of $\U(k_1/k_2)$--symmetric matrices. It was calculated in Refs. \onlinecite{Guh91} and \onlinecite{Guh96b} for a Wick rotation with angle $\psi=\pi/2$
\begin{equation}\label{4.4}
 \begin{split}
  \Delta_{s}^{(2,2;k_1k_2)}& = \frac{1}{B_{k_1k_2}^{(2,2)}(s_1,e^{\imath\psi}s_2)}\times\\
  & \times\left(\sum\limits_{j=1}^{k_1}\frac{\partial}{\partial s_{j1}}B_{k_1k_2}^{(2,2)}(s_1,e^{\imath\psi}s_2)\frac{\partial}{\partial s_{j1}}-e^{-2\imath\psi}\sum\limits_{j=1}^{k_2}\frac{\partial}{\partial s_{j2}}B_{k_1k_2}^{(2,2)}(s_1,e^{\imath\psi}s_2)\frac{\partial}{\partial s_{j2}}\right),\\
  B_{k_1k_2}^{(2,2)}(s_1,e^{\imath\psi}s_2)& = \frac{\Delta_{k_1}^2(s_1)\Delta_{k_2}^2(e^{\imath\psi} s_2)}{V_{k_1k_2}^2(s_1,e^{\imath\psi} s_2)} ,
 \end{split} 
\end{equation}
where we defined the mixed Vandermonde as $V_{k_1k_2}(s_1,e^{\imath\psi} s_2)=\prod\limits_{n=1}^{k_1}\prod\limits_{m=1}^{k_2}(s_{n1}-e^{\imath\psi} s_{m2})$ and $B_{k_1k_2}^{(2,2)}(s_1,e^{\imath\psi}s_2)$ is the Berezinian of the transformation from Cartesian to eigenvalue--angle coordinates.\cite{Guh91,Guh96,Guh96b} $s_{n1}$, $1\leq n\leq k_1$, and $s_{m2}$, $1\leq m \leq k_2$, are the eigenvalue bodies of the Hermitian matrix $\sigma_1$ in the boson--boson block and of the Hermitian matrix $\sigma_2$  in the fermion--fermion block, respectively. The operator $D_{s}^{(2,2;k_1k_2)}$ can be cast into a simpler form. Using the identities
\begin{eqnarray}
 \Delta_{s}^{(2;k)} & = & \frac{1}{\Delta_k(s)}\sum\limits_{j=1}^{k}\frac{\partial^2}{\partial s_{j}^2}\Delta_k(s), \label{4.5}\\
 \Delta_{s}^{(2,2;k_1k_2)} & = & \frac{1}{\sqrt{B_{k_1k_2}^{(2,2)}(s_1,e^{\imath\psi}s_2)}}\Str\frac{\partial^2}{\partial s^2}\sqrt{B_{k_1k_2}^{(2,2)}(s_1,e^{\imath\psi}s_2)} \label{4.6},
\end{eqnarray}
we find
\begin{equation}\label{4.7}
 \begin{split}
  D_{s}^{(2,2;k_1k_2)}& = \frac{e^{\imath\psi k_1k_2}}{(k_1k_2)!(4\pi)^{k_1k_2}}\frac{1}{\Delta_{k_1}(s_1)\Delta_{k_2}(e^{\imath\psi}s_2)}\times \\
  & \times\sum\limits_{n=0}^{k_1k_2}\binom{k_1k_2}{n}\left(\Str\frac{\partial^2}{\partial s^2}\right)^{k_1k_2-n}V_{k_1k_2}(s_1,e^{\imath\psi}s_2)\left(-\Str\frac{\partial^2}{\partial s^2}\right)^n\sqrt{B_{k_1k_2}^{(2,2)}(s_1,e^{\imath\psi} s_2)},
 \end{split}
\end{equation}
where we defined
\begin{equation}
\label{supertrace1}
 \Str\frac{\partial^2}{\partial s^2}=\sum\limits_{j=1}^{k_1}\frac{\partial^2}{\partial s_{j1}^2}-e^{-2\imath\psi}\sum\limits_{j=1}^{k_2}\frac{\partial^2}{\partial s_{j2}^2}\ .
\end{equation}
We obtain for the special case $k_1=k_2=1$ the well known result \cite{Con88,ConGro89,VWZ85}
\begin{equation}\label{4.8}
 D_{s}^{(2,2;1,1)}=\frac{e^{\imath\psi}}{2\pi}\frac{1}{s_1-e^{\imath\psi}s_2}\left(\frac{\partial}{\partial s_{1}}+e^{-\imath\psi}\frac{\partial}{\partial s_{2}}\right)\ .
\end{equation}

We now state an integral theorem, which is a generalization of the integral theorem due to Wegner worked out in Theorem 4.1 of Ref. \onlinecite{ConGro89} to a generalized Wick rotation and to an arbitrary dimension of the supermatrix.
\begin{theorem}[$\U(k_1/k_2)$--symmetric matrices]\label{t5}\ \\
 Let $f$ be a differentiable function of $\U(k_1/k_2)$--symmetric supermatrices of the form \eqref{4.1}, which is invariant under the action of $\U(k_1/k_2)$ and which has zero boundary condition at infinity, then
 \begin{equation}\label{4.9}
  \begin{split}
   & \int\limits_{\Herm(2,k_1)}\int\limits_{\Herm(2,k_2)} \int f(\sigma)d[\eta]d[\sigma_2]d[\sigma_1]=\\
  = & \begin{cases}
   \displaystyle(-1)^{k_1k_2}2^{-k_2(k_2-1)}\imath^{k_2^2}\left(\frac{e^{\imath\psi}}{2}\right)^{k_2(k_1-k_2)}\int
 \limits_{\Herm(2,k_1-k_2)}f(\tilde{\sigma}_1)d[\tilde{\sigma}_1] & , k_1>k_2\\
   \displaystyle2^{-k(k-1)}(-\imath)^{k^2}f(0) & , k_1=k_2=k\\
   \displaystyle2^{-k_1(k_1-1)}(-\imath)^{k_1^2}\left(\frac{e^{-\imath\psi}}{2}\right)^{k_1(k_2-k_1)}
 \int\limits_{\Herm(2,k_2-k_1)}f(e^{\imath\psi}\tilde{\sigma}_2)d[\tilde{\sigma}_2] & , k_1<k_2
  \end{cases}
  \end{split}
 \end{equation}
 where $d[\sigma_j]=\prod\limits_{n=1}^{k_j}d\sigma_{nnj}\prod\limits_{1\leq n< m\leq k_j}d{\rm Re}\ \sigma_{nmj}d{\rm Im}\ \sigma_{nmj}$ and $\Herm(2,N)$ denotes the space of ordinary Hermitian $N\times N$--matrices. $\tilde{\sigma}_1$ refers to the canonical embedding in the boson--boson matrix block and $\tilde{\sigma}_2$ refers to the canonical embedding in the fermion--fermion matrix block.
\end{theorem}
\textbf{Proof:}\ \\
First we prove the case $k_1=k_2=1$. We define the complex number $z=s_1-e^{\imath\psi}s_2$. Setting $f(s_1,s_2)=\tilde{f}(z,z^*)$ we get
\begin{equation}\label{4.10}
 \int\limits_{\Herm(2,1)}\int\limits_{\Herm(2,1)} \int f(\sigma)d[\eta]d[\sigma_2]d[\sigma_1]=\frac{1}{2\pi}\int\limits_{\mathbb{C}}\frac{1}{z}\frac{\partial}{\partial z^*}\tilde{f}(z,z^*)dzdz^*=-\imath \tilde{f}(0)=-\imath f(0).
\end{equation}
For arbitrary $k=k_1=k_2$ we rearrange and split the matrix $\sigma$ in the following way
\begin{equation}\label{4.11}
 \sigma=\begin{bmatrix} \sigma_s & v \\ v^{\dagger} & \tilde{\sigma} \end{bmatrix}  ,
\end{equation}
where $\sigma_s$ is a $\U(1/1)$--symmetric supermatrix and $\tilde{\sigma}$ a $\U(k-1/k-1)$--symmetric supermatrix. Moreover, we defined $v=(v_1,\hdots v_{k-1},w_1,\hdots w_{k-1})$ with the complex $(1/1)$--supervectors $v_{j}=\begin{bmatrix} z_j \\ e^{\imath\psi/2}\eta_j \end{bmatrix}$ and $w_{j}=\begin{bmatrix} e^{\imath\psi/2}\tilde{\eta}_j^* \\ e^{\imath\psi}\tilde{z}_j \end{bmatrix}$. The supervectors have the same structure as those of Theorem \ref{t3}. We integrate first over all variables except $\sigma_s$. The resulting function on the set of $\U(1/1)$--symmetric supermatrices
\begin{equation}\label{p4.1}
 F_1(\sigma_s)= \int\limits_{\Herm(2,k-1)}\int\limits_{\Herm(2,k-1)}\int\limits_{\mathbb{C}^{2(k-1)}} \int f(\sigma)d[v]d[\tilde{\sigma}]
\end{equation}
is invariant under the action of $\U(1/1)$. Therefore, we can use Eq. \eqref{4.10} and have to calculate
\begin{equation}\label{p4.2}
 \int\limits_{\Herm(2,k-1)}\int\limits_{\Herm(2,k-1)}\int\limits_{\mathbb{C}^{2k-3}} \int f\left(\begin{bmatrix} 0 & v \\ v^{\dagger} & \tilde{\sigma} \end{bmatrix}\right)d[v]d[\tilde{\sigma}].
\end{equation}
 We integrate over the remaining variables except over one pair of the $(1/1)$--supervectors $(u,u^\dagger)$ $=$ $(v_j,v_j^{\dagger})$ or $(u,u^\dagger)$ $=$ $(w_j,w_j^{\dagger})$. The function
\begin{equation}\label{p4.3}
 F_2(u,u^\dagger)= \int\limits_{\Herm(2,k_1-1)}\int\limits_{\Herm(2,k_2-1)}\int\limits_{\mathbb{C}^{2k-3}} \int f\left(\begin{bmatrix} 0 & v \\ v^{\dagger} & \tilde{\sigma} \end{bmatrix}\right)d[v_{\neq u}]d[\tilde{\sigma}]
\end{equation}
 fulfills the requirements of Theorem \ref{t3}. We perform the integration over all pairs $(v_j,v_j^{\dagger})$ and
 $(w_j,w_j^{\dagger})$ accordingly. We are left with an integration over $f(\tilde{\sigma})$. Since $\tilde{\sigma}$ has the 
same symmetry as $\sigma$ with lower matrix dimension, we can proceed by induction. This proves the second equation of Eq.~(\ref{4.9}). In order to prove the first and the third equation of Eq.~\eqref{4.9},  we define the $\U(k/k)$--symmetric supermatrix $\sigma_{k}$, where $k= {\rm min}(k_1,k_2)$, and the $\Delta k\times\Delta k$--Hermitian matrix $\sigma_{\Delta k}$ in the boson--boson ($k=k_2$) or in the fermion--fermion ($k=k_1$) block, where $\Delta k= |k_1-k_2|$. We define the function
 \begin{equation}\label{p4.4}
  F_3(\sigma_{k})= \int\limits_{\Herm(2,\Delta k)}\int\limits_{\left(\mathbb{C}^{k}\right)^{\Delta k}} 
   \int f\left(\begin{bmatrix} \sigma_{k} & v \\ v^{\dagger} & \sigma_{\Delta k} \end{bmatrix}\right)d[v]d[\sigma_{\Delta k}]
 \end{equation}
 and apply the second equation of Eq.~(\ref{4.9}) on $F_3$. The off-diagonal block matrix $v$ consists of complex $(k/k)$--supervectors. We iteratively perform the integrations over these supervectors using Theorem \ref{t3}. This completes the proof.\hfill$\square$

We next consider supermatrices which are form invariant under the action of $\UOSp^{(+)}(k_1/2k_2)$ or $\UOSp^{(-)}(2k_1/k_2)$. We first focus on the representation $\UOSp^{(+)}(k_1/2k_2)$ of the supergroup $\UOSp(k_1/2k_2)$ and later extend our results to $\UOSp^{(-)}(2k_2/k_1)$. $\UOSp^{(+)}(k_1/2k_2)$--symmetric matrices have the form
\begin{equation}\label{4.12}
 \Sigma=\begin{bmatrix} \Sigma_1 & e^{\imath\psi/2}\alpha & e^{\imath\psi/2}\alpha^* \\ -e^{\imath\psi/2}\alpha^{\dagger} & e^{\imath\psi}\Sigma_{21} & e^{\imath\psi}\Sigma_{22} \\ e^{\imath\psi/2}\alpha^T & -e^{\imath\psi}\Sigma_{22}^* & e^{\imath\psi}\Sigma_{21}^* \end{bmatrix}
\end{equation}
where the body of $\Sigma_1$ is a $k_1\times k_1$--real symmetric matrix, the body of $\Sigma_{21}$ is a $k_2\times k_2$--Hermitian matrix, and the body of $\Sigma_{22}$ is a $k_2\times k_2$--complex antisymmetric matrix. Therefore, the body of the fermion--fermion block is a $2k_2\times 2k_2$--quaternionic self-adjoint matrix. $\alpha$ is a $k_1\times k_2$--matrix with independent anticommuting entries. The generalized Wick rotation is introduced for the same reason as before.

Let $f$ be an invariant function on the space of supermatrices of the form (\ref{4.12})
\begin{equation}
\label{4.12a}
f(\Sigma) = f(U^{-1}\Sigma U) \ ,\ \quad U \in  \UOSp(k_1/k_2) \ .
\end{equation}
The radial part of the space of supermatrices of the form (\ref{4.12}) is the space of diagonal matrices $s=\diag(s_1,e^{\imath\psi}s_2)$ and  
\begin{equation}\label{h4.5}
s_1 = \diag(s_{11},\ldots,s_{k_11}) \ ,\ s_2\ =\ \diag(s_{12},\ldots,s_{k_22},s_{12},\ldots,s_{k_22})\ .
\end{equation}
The metric is
\begin{equation}\label{h4.2}
 g_n=\begin{cases}
      	1	&	,\ n\text{ is a diagonal index in the boson--boson block,}\\
	2	&	,\ n\text{ is an off-diagonal index in the boson--boson block or}\\
		&	\text{ a diagonal index in the fermion--fermion block,}\\
	4	&	,\ n\text{ is an off-diagonal index in the fermion--fermion block}
     \end{cases}
\end{equation}
and $h_m=2\imath$. Applying Theorem \ref{t1} yields for the integration over the Grassmann variables of an invariant superfunction $f$
\begin{eqnarray}\label{4.13}
 \int f(\sigma)d[\eta] &=&  \frac{e^{\imath\psi k_1k_2}}{(k_1k_2)!(4\pi)^{k_1k_2}}
\sum\limits_{n=0}^{k_1k_2}\binom{k_1k_2}{n}\left(\Delta_{s_1}^{(1;k_1)}-e^{-2\imath\psi}\Delta_{s_2}^{(4;k_2)}\right)^{k_1k_2-n}
    \left(-\Delta_{s}^{(1,4;k_1k_2)}\right)^nf(s)= \nonumber\\
 &=&D_{s}^{(1,4;k_1k_2)}f(s)  .
\end{eqnarray}
Here, we used the radial parts of the Laplacians in the space of symmetric matrices $\Delta_{s}^{(1;k)}$ and in the space of quaternionic self-adjoint matrices  $\Delta_{s}^{(4;k)}$ 
\begin{equation}\label{4.15}
 \Delta_{s}^{(1;k)} = \sum\limits_{j=1}^{k}\frac{1}{\Delta_k(s)}\frac{\partial}{\partial s_j}\Delta_k(s)\frac{\partial}{\partial s_j}\ \ {\rm and}\ \ \Delta_{s}^{(4;k)} = \frac{1}{2}\sum\limits_{j=1}^{k}\frac{1}{\Delta_k^4(s)}\frac{\partial}{\partial s_j}\Delta_k^4(s)\frac{\partial}{\partial s_j} .
\end{equation}
The radial part of the Laplacian in the superspace of $\UOSp^{(+)}(k_1/2k_2)$--symmetric supermatrices (\ref{h4.5}) reads
\begin{equation}\label{4.16}
 \begin{split}
  \Delta_{s}^{(1,4;k_1k_2)}&= \frac{1}{B_{k_1k_2}^{(1,4)}(s_1,e^{\imath\psi}s_2)}\times\\
  & \times\left(\sum\limits_{j=1}^{k_1}\frac{\partial}{\partial s_{j1}}B_{k_1k_2}^{(1,4)}(s_1,e^{\imath\psi}s_2)\frac{\partial}{\partial s_{j1}}-\frac{e^{-2\imath\psi}}{2}\sum\limits_{j=1}^{k_2}\frac{\partial}{\partial s_{j2}}B_{k_1k_2}^{(1,4)}(s_1,e^{\imath\psi}s_2)\frac{\partial}{\partial s_{j2}}\right),\\
  B_{k_1k_2}^{(1,4)}(s_1,e^{\imath\psi}s_2)&= \frac{\Delta_{k_1}(s_1)\Delta_{k_2}^4(e^{\imath\psi} s_2)}{V_{k_1k_2}^2(s_1,e^{\imath\psi} s_2)} ,
 \end{split} 
\end{equation}
see Ref. \onlinecite{Guh96b}. As in the $\U(k_1/k_2)$ case, we can simplify $D_{s}^{(1,4;k_1k_2)}$ using the identities
\begin{eqnarray}
 \Delta_{s}^{(1;k)} & = & \frac{1}{\sqrt{|\Delta_k(s)|}}H_{s}^{(1;k)}\sqrt{|\Delta_k(s)|}\ , \label{4.17}\\
 \Delta_{s}^{(4;k)} & = & \frac{1}{\Delta_k^2(s)}H_{s}^{(4;k)}\Delta_k^2(s)\ \ {\rm and} \label{4.18}\\
 \Delta_{s}^{(1,4;k_1k_2)} & = & \frac{1}{\sqrt{B_{k_1k_2}^{(1,4)}(s_1,e^{\imath\psi}s_2)}}H_{s}^{(1,4;k_1k_2)}\sqrt{B_{k_1k_2}^{(1,4)}(s_1,e^{\imath\psi}s_2)} , \label{4.19}
\end{eqnarray}
where we introduced the operators 
\begin{eqnarray}
 H_{s}^{(1;k)} & = & \sum\limits_{j=1}^{k}\frac{\partial^2}{\partial s_{j}^2}+\frac{1}{2}\sum\limits_{1\leq m<n\leq k}\frac{1}{(s_n-s_m)^2} \label{4.20} ,\\
 H_{s}^{(4;k)} & = & \frac{1}{2}\sum\limits_{j=1}^{k}\frac{\partial^2}{\partial s_{j}^2}-2\sum\limits_{1\leq m<n\leq k}\frac{1}{(s_n-s_m)^2} \label{4.21} ,\\
 H_{s}^{(1,4;k_1k_2)} & = & H_{s_1}^{(1;k_1)}-e^{-2\imath\psi}H_{s_2}^{(4;k_2)}-\sum\limits_{m=1}^{k_1} \sum\limits_{n=1}^{k_2}\frac{1}{(s_{m1}-e^{\imath\psi} s_{n2})^2} \label{4.22} .
\end{eqnarray}
As in the $\U(k_1/k_2)$ case, this transformation is useful because the Laplacians are represented in a Hamiltonian form. The ordinary matrix Bessel functions times the square root of the Vandermonde determinant and the supermatrix Bessel functions times the square root of the Berezinian are eigenfunctions of $H_{r}^{(\beta;k)}$ and $H_{r}^{(\beta,4/\beta;k_1k_2)}$, respectively. One can find the definition of the matrix Bessel functions in Refs. \onlinecite{GuhKoh02} and \onlinecite{GuhKoh02b}, see also Eq. \eqref{5.1}. We hope that we can calculate an explicit formula for supermatrix Bessel functions depending on matrix Bessel functions in the ordinary space. The representation above might be useful for this purpose. In Sec.~\ref{application}, we will show that this is indeed a very helpful tool to calculate the supermatrix Bessel function for the $\U(k_1/k_2)$ case.

One obtains
\begin{equation}\label{4.23}
 \begin{split}
	D_{s}^{(1,4;k_1k_2)} & = \frac{e^{\imath\psi k_1k_2}}{(k_1k_2)!(4\pi)^{k_1k_2}}\frac{1}{\sqrt{\Delta_{k_1}(s_1)}\Delta_{k_2}^2(e^{\imath\psi}s_2)}\sum\limits_{n=0}^{k_1k_2}\binom{k_1k_2}{n}\times\\
        & \times
        \left(H_{s_1}^{(1;k_1)}-e^{-2\imath\psi}H_{s_2}^{(4;k_2)}\right)^{k_1k_2-n}V_{k_1k_2}(s_1,e^{\imath\psi}s_2)\left(-H_{s}^{(1,4;k_1k_2)}\right)^n\sqrt{B_{k_1k_2}^{(1,4)}(s_1,e^{\imath\psi}s_2)} .
 \end{split}
\end{equation}

We give the two simplest examples for $k_2=1$ for illustrative purposes. For $k_1=1$, we have 
\begin{equation}\label{4.24}
 D_{s}^{(1,4;1,1)}=\frac{e^{\imath\psi}}{4\pi}\frac{1}{s_1-e^{\imath\psi}s_2}\left(2\frac{\partial}{\partial s_1}+e^{-\imath\psi}\frac{\partial}{\partial s_2}\right) ,
\end{equation}
and for $k_1=2$, we find
\begin{equation}\label{4.25}
 \begin{split}
  D_{s}^{(1,4;2,1)} & =\frac{e^{2\imath\psi}}{4\pi^2}\frac{1}{s_{11}-s_{21}}\left[\left(2\frac{\partial}{\partial s_{11}}+e^{-\imath\psi}\frac{\partial}{\partial s_2}\right)\frac{1}{s_{21}-e^{\imath\psi}s_2}\left(2\frac{\partial}{\partial s_{21}}+e^{-\imath\psi}\frac{\partial}{\partial s_2}\right)-\right.\\
  &\left.-\left(2\frac{\partial}{\partial s_{21}}+e^{-\imath\psi}\frac{\partial}{\partial s_2}\right)\frac{1}{s_{11}-e^{\imath\psi}s_2}\left(2\frac{\partial}{\partial s_{11}}+e^{-\imath\psi}\frac{\partial}{\partial s_2}\right)\right] .
 \end{split}
\end{equation}
The second example is needed to prove the following integral theorem.
\begin{theorem}[$\UOSp^{(+)}(k_1/2k_2)$--symmetric matrices]\label{t6}\ \\
 Let $f$ be a differentiable function on $\UOSp^{(+)}(k_1/2k_2)$--symmetric supermatrices of the form (\ref{4.12}), which is invariant under the action of $\UOSp^{(+)}(k_1/2k_2)$ and which has zero boundary condition at infinity. In addition it fulfills the condition 
\begin{equation}\label{t6r}
 \left.\left(2\frac{\partial}{\partial s_{n1}}+e^{-\imath\psi}\frac{\partial}{\partial s_{m2}}\right)f(s)\right|_{s_{n1}=s_{m2}=0}=0
\end{equation}
for all pairs of eigenvalues of the $\UOSp^{(+)}(k_1/2k_2)$--symmetric supermatrices, then
 \begin{equation}\label{4.26}
  \begin{split}
   & \int\limits_{\Herm(1,k_1)}\int\limits_{\Herm(4,k_2)} \int f(\sigma)d[\eta]d[\sigma_2]d[\sigma_1]=\\
  = & \begin{cases}
	\displaystyle\left(2\imath e^{\imath\psi}\right)^{k_2} 2^{-k_2(k_2-1)}(-e^{\imath\psi})^{k_2(k_1-2k_2)}\int\limits_{\Herm(1,k_1-2k_2)}f(\tilde{\sigma}_1)d[\tilde{\sigma}_1] & ,\ k_1>2k_2\\
	& \\
	\displaystyle\left(2\imath e^{\imath\psi}\right)^k 2^{-k(k-1)}f(0) & ,\ k_1/2=k_2=k\\
	& \\
	\displaystyle\left(2\imath e^{\imath\psi}\right)^{k_1/2} 2^{-k_1(k_1/2-1)/2}\left(\frac{e^{-\imath\psi}}{2}\right)^{k_1(k_2-k_1/2)}\times & ,\ k_1<2k_2\ \wedge\\ \displaystyle\times\int\limits_{\Herm(4,k_2-k_1/2)}f(e^{\imath\psi}\tilde{\sigma}_2)d[\tilde{\sigma}_2] & \ \ k_1\in(2\mathbb{N}_0)\\
	& \\
	\displaystyle\left(-2\imath e^{2\imath\psi}\right)^{(k_1-1)/2} 2^{-(k_1-1)(k_1-3)/4}\left(\frac{e^{-\imath\psi}}{2}\right)^{(k_1-1)(k_2-(k_1-1)/2)}\times & ,\ k_1<2k_2\ \wedge\\ 
	\times\int\limits_{\mathbb{R}}\int\limits_{\Herm(4,k_2-(k_1-1)/2)}\int f\left(\begin{bmatrix} \tilde{\sigma}_1 & e^{\imath\psi/2}\tilde{\eta}^{\dagger} \\ e^{\imath\psi/2}\tilde{\eta} & e^{\imath\psi}\tilde{\sigma}_2 \end{bmatrix}\right)d[\tilde{\eta}]d[\tilde{\sigma}_2]d\tilde{\sigma}_1 &  \ \ k_1\in(2\mathbb{N}_0+1)
      \end{cases}
  \end{split}
 \end{equation}
where
\begin{eqnarray*}
 d[\sigma_1] & = & \prod\limits_{1\leq n\leq m\leq k_1}d\sigma_{nm1} ,\\
 d[\sigma_2] & = & \prod\limits_{n=1}^{k_2}d\sigma_{nn21}\prod\limits_{1\leq n< m\leq k_2}d{\rm Re}\ \sigma_{nm21}d{\rm Im}\ \sigma_{nm21}d{\rm Re}\ \sigma_{nm22}d{\rm Im}\ \sigma_{nm22}
\end{eqnarray*}
and where $\Herm(1,N)$ is the space of real symmetric $N\times N$--matrices and $\Herm(4,N)$ is the space of quaternionic self-adjoint $2N\times 2N$--matrices. $\tilde{\sigma}_1$ refers to the canonical embedding in the boson--boson matrix block, $\tilde{\sigma}_2$ refers to the canonical embedding in the fermion--fermion matrix block and $\tilde{\eta}$ refers to the canonical embedding in the boson--fermion matrix block.
\end{theorem}
The idea of the proof is to apply the recursive method of Wegner\cite{Weg83,ConGro89} using the operator $D_s^{(1,4;k_1k_2)}$.\\
\textbf{Proof:}\ \\
As in Theorem \ref{t5}, we first prove the simplest nontrivial case $k_1=2k_2=2$ . We use \eqref{4.25} and obtain after an integration over eigenvalue and angular coordinates
\begin{equation}\label{4.27}
 \begin{split}
  & \int\limits_{\Herm(1,2)}\int\limits_{\Herm(4,1)} \int f(\sigma)d[\eta]d[\sigma_2]d[\sigma_1]=\int\limits_{\mathbb{R}^2}\int\limits_{\mathbb{R}}\frac{e^{2\imath\psi}}{2\pi}\frac{|s_{11}-s_{21}|}{s_{11}-s_{21}}\times\\
  &\times\left(2\frac{\partial}{\partial s_{11}}+e^{-\imath\psi}\frac{\partial}{\partial s_2}\right)\frac{1}{s_{21}-e^{\imath\psi}s_2}\left(2\frac{\partial}{\partial s_{21}}+e^{-\imath\psi}\frac{\partial}{\partial s_2}\right)f(s_{11},s_{21},s_2)ds_{2}d[s_1] .
 \end{split}
\end{equation}
The equation is valid because the single terms in the integrand, see Eq. \eqref{4.25}, are symmetric under interchange of the two bosonic eigenvalues. A change in variables $r=\frac{1}{2}(s_{11}-s_{21})$ and $z=\frac{1}{2}(s_{11}+s_{21})-e^{\imath\psi}s_2$, such that $f(s_{11},s_{21},s_2)=\tilde{f}(r,z,z^*)$, leads to
\begin{equation}\label{4.28}
 \begin{split}
  & \int\limits_{\Herm(1,2)}\int\limits_{\Herm(4,1)} \int f(\sigma)d[\eta]d[\sigma_2]d[\sigma_1]=\\
  = &  C_\psi\int\limits_{\mathbb{C}}\int\limits_{\mathbb{R}^+}\left(\left(1-e^{-2\imath\psi}\right)\frac{\partial}{\partial z^*}+\frac{\partial}{\partial r}\right)\frac{1}{z-r}\left(\left(1-e^{-2\imath\psi}\right)\frac{\partial}{\partial z^*}-\frac{\partial}{\partial r}\right)\tilde{f}(r,z,z^*)drdzdz^*=\\
  \overset{(2)}{=} & -C_\psi \int\limits_{\mathbb{C}}\left.\frac{1}{z}\left(\left(1-e^{-2\imath\psi}\right)\frac{\partial}{\partial z^*}-\frac{\partial}{\partial r}\right)\tilde{f}(r,z,z^*)\right|_{r=0}dzdz^*=\\
  \overset{(3)}{=} & -\frac{e^{\imath\psi}}{\pi}\int\limits_{\mathbb{C}}\frac{1}{z}\frac{\partial}{\partial z^*}\tilde{f}(0,z,z^*)dzdz^*=2\imath e^{\imath\psi}\tilde{f}(0)=2\imath e^{\imath\psi}f(0)
 \end{split}
\end{equation}
where $C_\psi=e^{2\imath\psi}/2\pi\imath\sin(\psi)$. The second equality (2) holds because the integral over the complex plane with the derivative with respect to $z^*$ is up to a constant equal to
\begin{equation}\label{p4.0}
\begin{split}
 & \int\limits_{\mathbb{R}^+}\left.\left(\left(1-e^{-2\imath\psi}\right)\frac{\partial}{\partial z^*}-\frac{\partial}{\partial r}\right)\tilde{f}(r,z,z^*)\right|_{z=z^*=r}dr=\\
 = & \int\limits_{\mathbb{R}^+}\left.\left(2\frac{\partial}{\partial s_{21}}+e^{-\imath\psi}\frac{\partial}{\partial s_2}\right)f(s_{11},s_{21},s_2)\right|_{s_{21}=s_{2}=0}ds_{11}\overset{\eqref{t6r}}{=}0 .
\end{split}
\end{equation}
The third equality (3) holds because of the symmetry $f(s_{11},s_{21},s_2)=f(s_{21},s_{11},s_2)$ and accordingly we get $\tilde{f}(r,z,z^*)=\tilde{f}(-r,z,z^*)$ and $\partial\tilde{f}/\partial r(r,z,z^*)|_{r=0}=0$ .

For arbitrary $k=k_1=k_2/2$ we proceed as in the proof of Theorem \ref{t5}. We split off a $\UOSp^{(+)}(2/2)$--symmetric supermatrix and integrate over the remaining variables such that the resulting function on the set of $\UOSp^{(+)}(2/2)$--symmetric supermatrices is invariant. We apply the simplest case of the theorem above. We iteratively perform the integrals over the $2(k-1)$ real $(2/2)$--supervectors and $(k-1)$ quaternionic $(2/2)$--supervectors in the off-diagonal matrix block in the same manner as in the $\U(k/k)$ case with help of the Theorems \ref{t2} and \ref{t4}. Finally we carry out iteratively the integral over a $\UOSp^{(+)}(2k-2/2k-2)$--symmetric supermatrix. This proves the second equation in Eq.~(\ref{4.26}).

For arbitrary $k_1$ and $k_2$, one can split off the largest $\UOSp^{(+)}(2k/2k)$--symmetric supermatrix where $k$ is the minimum of $k_2$ and $(k_1-k_1{\rm mod}2)/2$. We use the second equation of Eq.~(\ref{4.26}) to treat this block. The integrations over the $(2k/2k)$--supervectors, which are $(k_1-2k_2)$ real supervectors for $k_1>2k_2$ and $(2k_2-k_1+k_1{\rm mod}2)/2$ quaternionic supervectors for $k_1<2k_2$ plus an additional real supervector depending on whether $k_1$ is even or odd, can be iteratively calculated with help of the Theorems in Sec. III. This proves the first, the third and the fourth equation of Eq.~(\ref{4.26}). \hfill$\square$\\

We remark that the property \eqref{t6r} is not a strong restriction on the set of functions. For example, the class of functions which is $C^{\infty}$-differentiable in their supertraces satisfies this condition. Furthermore, we remark here that there is no such integral theorem for  $\UOSp(1/2)$--symmetric supermatrices and thus there is no integral reduction as above for $\UOSp(1/2k_2)$, see Appendix A.

The extension of these results to invariant functions of $\UOSp^{(-)}(2k_1/k_2)$--symmetric supermatrices is straightforward. The metric is
\begin{equation}\label{h4.4}
 g_n=\begin{cases}
      	1	&	,\ n\text{ is a diagonal index in the fermion--fermion block,}\\
	2	&	,\ n\text{ is an off-diagonal index in the fermion--fermion block or}\\
		&	\text{ a diagonal index in the boson--boson block,}\\
	4	&	,\ n\text{ is an off-diagonal index in the boson--boson block}
     \end{cases}
\end{equation}
and $h_m=2\imath$. The differential operator $D_{s}^{(4,1;k_1k_2)}$ obeys the following symmetry relation 
\begin{equation}\label{4.29}
D_{s}^{(4,1;k_1k_2)}f(s) \ =\ (-1)^{k_1k_2}D_{\tilde{s}}^{(1,4;k_2k_1)}f(s)
\end{equation}
where $\tilde{s}$ is related to $s$ by $\tilde{s}=\diag(-e^{\imath\psi}s_2,-s_1)$.
The corresponding theorem follows directly from Theorem \ref{t6}.
\begin{theorem}[$\UOSp^{(-)}(2k_1/k_2)$--symmetric matrices]\label{t7}\ \\
Let the measures be the same as in Theorem \ref{t6}. Let $f$ be a differentiable function on $\UOSp^{(-)}(k_1/2k_2)$--symmetric supermatrices, which is invariant under the action of $\UOSp^{(-)}(k_1/2k_2)$ and which has zero boundary condition at infinity. In addition it fulfills the condition 
\begin{equation}\label{t7r}
 \left.\left(\frac{\partial}{\partial s_{n1}}+2e^{-\imath\psi}\frac{\partial}{\partial s_{m2}}\right)f(s)\right|_{s_{n1}=s_{m2}=0}=0
\end{equation}
for all pairs of eigenvalues of the $\UOSp^{(-)}(2k_1/k_2)$--symmetric matrices, then\\
 \begin{equation}\label{4.30}
  \begin{split}
   & \int\limits_{\Herm(4,k_1)}\int\limits_{\Herm(1,k_2)} \int f(\sigma)d[\eta]d[\sigma_2]d[\sigma_1]=\\
  = & \begin{cases}
	\displaystyle\left(-2\imath e^{-\imath\psi}\right)^{k_2/2} 2^{-k_2(k_2/2-1)/2}\left(\frac{e^{\imath\psi}}{2}\right)^{k_2(k_1-k_2/2)}\times & ,\ 2k_1>k_2\ \wedge\\ \displaystyle\times\int\limits_{\Herm(4,k_1-k_2/2)}f(\tilde{\sigma}_1)d[\tilde{\sigma}_1] & \ \ k_2\in(2\mathbb{N}_0)\\
	& \\
	\displaystyle\left(2\imath e^{-2\imath\psi}\right)^{(k_2-1)/2} 2^{-(k_2-1)(k_2-3)/4}\left(\frac{e^{\imath\psi}}{2}\right)^{(k_2-1)(k_1-(k_2-1)/2)}\times & ,\ 2k_1>k_2\ \wedge\\
	\displaystyle\times \int\limits_{\mathbb{R}}\int\limits_{\Herm(4,k_1-(k_2-1)/2)}\int f\left(\begin{bmatrix} \tilde{\sigma}_1 & e^{\imath\psi/2}\tilde{\eta}^{\dagger} \\ e^{\imath\psi/2}\tilde{\eta} & e^{\imath\psi}\tilde{\sigma}_2 \end{bmatrix}\right)d[\tilde{\eta}]d\tilde{\sigma}_2d[\tilde{\sigma}_1] & \ \ k_2\in(2\mathbb{N}_0+1)\\
	& \\
	\displaystyle\left(-2\imath e^{-\imath\psi}\right)^k 2^{-k(k-1)}f(0) & ,\ k_1=\frac{k_2}{2}=k\\
	& \\
	\displaystyle\left(-2\imath e^{-\imath\psi}\right)^{k_1} 2^{-k_1(k_1-1)}e^{-\imath\psi k_1(k_2-2k_1)}\int\limits_{\Herm(1,k_2-2k_1)}f(e^{\imath\psi}\tilde{\sigma}_2)d[\tilde{\sigma}_2] & ,\ 2k_1<k_2
      \end{cases}
  \end{split}
 \end{equation}
where $\tilde{\sigma}_1$, $\tilde{\sigma}_2$ and $\tilde{\eta}$ have the same meaning as in Theorem \ref{t6}.
\end{theorem}

We now investigate the structure of the differential operators $D_{s}^{(\beta,4/\beta;k_1k_2)}$, where we introduced $\beta\in\{1,2,4\}$ usually called the Dyson index. The Laplacians in ordinary space  equations (\ref{4.3}) and (\ref{4.15}), respectively, in superspace equations (\ref{4.4}) and (\ref{4.16}), can be written as
\begin{eqnarray}\label{4.31}
 \Delta_{s}^{(\beta;k)} &= &\gamma(\beta)\left(\sum\limits_{n=1}^k\frac{\partial^2}{\partial s_n^2}+\sum\limits_{1\leq m<n\leq k}\frac{\beta}{s_m-s_n}\left(\frac{\partial}{\partial s_m}-\frac{\partial}{\partial s_n}\right)\right) ,\\
\Delta_{s}^{(\beta,4/\beta;k_1k_2)}
                 & = & \Delta_{s_1}^{(\beta;k_1)}-e^{-2\imath\psi}\Delta_{s_2}^{(4/\beta;k_2)}-\nonumber\\
       & - & \sum\limits_{\begin{subarray}\ 1\leq m\leq k_1 \\ 1\leq n\leq k_2 \end{subarray}} \frac{2}{s_{m1}-e^{\imath\psi}s_{n2}}\left(\gamma(\beta)
   \frac{\partial}{\partial s_{m1}}-\gamma\left(\frac{4}{\beta}\right)e^{-\imath\psi}\frac{\partial}{\partial s_{n2}}\right) \label{4.32} .
\end{eqnarray}
Here, $\gamma(\beta)$ is unity for $\beta\in\{1,2\}$ and 2 for $\beta=4$ case. We now introduce a set of operators $D^{(\mu,\nu)}(s_a,s_b)$,
\begin{eqnarray}
 D^{(\beta,\beta)}(s_{nj},s_{mj}) & = & \frac{1}{s_{nj}-s_{mj}}\left(\frac{\partial}{\partial s_{nj}}-\frac{\partial}{\partial s_{mj}}\right)\ ,\ j\in\{1,2\}, \label{4.34}\\
 D^{(\beta,4/\beta)}(s_{n1},s_{m2}) & = & \frac{1}{s_{n1}-e^{\imath\psi}s_{m2}}\left(\gamma(\beta)\frac{\partial}{\partial s_{n1}}-\gamma\left(\frac{4}{\beta}\right)e^{-\imath\psi}\frac{\partial}{\partial s_{m2}}\right) \label{4.35} .
\end{eqnarray}
We notice that
\begin{equation}\label{4.33}
 \left[\Str\frac{\partial^2}{\partial s^2},D^{(\mu,\nu)}(s_a,s_b)\right]_-=-2\left(D^{(\mu,\nu)}(s_a,s_b)\right)^2
\end{equation}
for all operators $D^{(\mu,\nu)}(s_a,s_b)$, where
\begin{equation}\label{4.36}
 \Str\frac{\partial^2}{\partial s^2}=\gamma(\beta)\sum\limits_{n=1}^{k_1}\frac{\partial^2}{\partial s_{n1}^2}-\gamma\left(\frac{4}{\beta}\right)e^{-2\imath\psi}\sum\limits_{n=1}^{k_2}\frac{\partial^2}{\partial s_{n2}^2}\ .
\end{equation}
Now, we recall Eq. \eqref{l2.1} and combine it with \eqref{4.33}. We see that $D_{s}^{(\beta,4/\beta;k_1,k_2)}$ is homogeneous in the operators $D^{(\beta,\beta)}(s_{nj},s_{mj})$ and $D^{(\beta,4/\beta)}(s_{n1},s_{m2})$ of degree $k_1k_2$.

\section{Applications}
\label{application}

In this section we give two examples for the usefulness of the formalism developed previously. It is well known in random matrix theory that the energy density $\rho(x)$ of a  Gaussian random matrix ensemble can be expressed as the derivative with respect a source term of a generating function $Z(x+J)$. This generating function has a representation as a matrix integral over certain spaces of supermatrices. For the Gaussian unitary ensemble (GUE), it is given by
 \begin{equation}\label{5.5}
 Z(x_1,J_1)=C\int\limits_{\rm Herm(2,1)}\int\limits_{\rm Herm(2,1)} \int e^{-\Str\left(\sigma+x+J\right)^2}\Sdet^{-N}(\sigma+\imath\epsilon\e_2)d[\eta]d[\sigma_2]d[\sigma_1],
\end{equation} 
where $\sigma$ is a $\U(1/1)$--symmetric supermatrix as defined in Eq.~(\ref{4.1}). $C$ is a normalization constant for $J=0$ and $x=x_1\e_2$ with $x_1,\epsilon\in\mathbb{R}$. $\Sdet$ is the superdeterminant, $N$ is the level number, and $J=\diag(-J_1,J_1)$ with $J_1\in\mathbb{R}$. 

For the Gaussian orthogonal ensemble (GOE) the generating function of the energy density is given by 
\begin{equation}\label{5.14}
 Z(x_1,J_1)=C\int\limits_{\Herm(1,2)}\int\limits_{\Herm(4,2)} \int e^{-\Str(\sigma+x+J)^2}\Sdet^{-N/2}(\sigma+\imath\epsilon\e_4)d[\eta]d[\sigma_2]d[\sigma_1],
\end{equation}
where $\sigma$ is a $\UOSp^{(+)}(2/2)$--symmetric supermatrix as defined in Eq.~(\ref{4.12}), $C$ is the normalization constant, and $x=x_1\e_4$. $N$ is the level number and $J=\diag(-J_1,-J_1,J_1,J_1)$ with $J_1\in\mathbb{R}$.

In our first example we show how these supermatrix integrals are efficiently evaluated within the present formalism.

The second example concerns the calculation of the supermatrix Bessel function, defined as the supersymmetric group integral,\cite{GuhKoh02,GuhKoh02b}
\begin{equation}\label{b1}
 \Phi_{k_1k_2}^{(2,2)}(s,x)=\int\limits_{\U(k_1/k_2)}e^{\Str(sUxU^{\dagger})}d\mu(U),
\end{equation}
where $d\mu(U)$ is the Haar--measure of the group $\U(k_1/k_2)$ and $s$ and $x$ are two diagonal $(k_1/k_2)$-supermatrices. We rederive within the present formalism the result, derived in Refs. \onlinecite{Guh91} and \onlinecite{Guh96}.

\subsection{One--point correlation functions and supermatrix Bessel functions}

We start with a GUE. We consider the integral
\begin{equation}\label{5.1}
 I(x_1,x_2)=\int\limits_{\Herm(2,1)}\int\limits_{\Herm(2,1)} \int f(\sigma)e^{\Str\sigma x}d[\eta]d[\sigma_2]d[\sigma_1]
\end{equation}
where $x={\rm diag}(x_1,x_2)$ is a diagonal $(1/1)$--supermatrix. Here, $f$ is a $\U(1/1)$--invariant superfunction on the $\U(1/1)$--symmetric supermatrices with zero boundary condition at infinity. Since Grassmannian variables are contained in $f$ only, we can apply Theorem \ref{t1} and find
\begin{equation}\label{5.2}
 I(x_1,x_2)=\int\limits_{\mathbb{R}}\int\limits_{\mathbb{R}} e^{\left(s_1 x_1-e^{\imath\psi}s_2x_2\right)} D_{s}^{(2,2;1,1)}f(s)ds_2ds_1 .
\end{equation}
Now, we perform an integration by parts and shift the differential operator onto the exponential function
\begin{equation}\label{5.3}
 I(x_1,x_2)=\int\limits_{\mathbb{R}}\int\limits_{\mathbb{R}}\left( D_{s}^{(2,2;1,1)}f(s)e^{\left(s_1 x_1-e^{\imath\psi}s_2x_2\right)}- f(s)D_{s}^{(2,2;1,1)} e^{\left(s_1 x_1-e^{\imath\psi}s_2x_2\right)}\right)ds_2ds_1 .
\end{equation}
Due to the simplicity of $D_{s}^{(2,2;1,1)}$, we apply the Cauchy--integral theorem and obtain
\begin{equation}\label{5.4}
 I(x_1,x_2)=-\imath f(0)-\frac{e^{\imath\psi}}{2\pi}\int\limits_{\mathbb{R}}\int\limits_{\mathbb{R}} f(s_1,s_2)\frac{x_1-x_2}{s_1-e^{\imath\psi}s_2}e^{\left(s_1 x_1-e^{\imath\psi}s_2x_2\right)}ds_2ds_1 .
\end{equation}
The integrand in the second term of $I(x)$ is a product of the invariant function $f$, the Berezinian $B_{1,1}^{(2,2)}(s_1,s_2)= (s_1-e^{\imath\psi}s_2)^{-2}$, and the supermatrix Bessel function which coincides with Ref. \onlinecite{Guh91}. We remark that \eqref{5.4} agrees with the known general transformation from the Cartesian coordinates to the eigenvalue--angle coordinates for Wick rotation $\imath$.\cite{Guh93b} 
Since the generating function $Z$ [see Eq.~(\ref{5.5})] is exactly of the form (\ref{5.1}), we can use Eq.~\eqref{5.4} and find
\begin{equation}\label{5.6}
 Z(x+J)=1+\frac{e^{\imath\psi}}{2\pi}C\int\limits_{\mathbb{R}}\int\limits_{\mathbb{R}}e^{-\Str(s+x+J)^2}\left(\frac{s_2+\imath\epsilon}{s_1+\imath\epsilon}\right)^{N}\frac{4J_1}{s_1-e^{\imath\psi}s_2}ds_2ds_1
\end{equation}
which is indeed the correct result.\cite{Guh91,Guh06} Furthermore, we identify the boundary term in \eqref{5.6} and \eqref{5.4} as the Efetov--Wegner term.\cite{Guh91,Guh06,BasAke07} This term guarantees the normalization of $Z$ at $J_1=0$.

In analogy to the GUE case, we consider the following integral for the GOE
\begin{equation}\label{5.7}
 I(x_{11},x_{21},x_2)=\int\limits_{\Herm(4,2)}\int\limits_{\Herm(1,1)} \int f(\sigma)e^{\Str\sigma x}d[\eta]d[\sigma_2]d[\sigma_1],
\end{equation}
where $x={\rm diag}(x_{11},x_{21},x_2,x_2)$ is a diagonal $(2/2)$--supermatrix. Now, $f$ is an $\UOSp^{(+)}(2/2)$--invariant superfunction on the space of $\UOSp^{(+)}(2/2)$--symmetric supermatrices with zero boundary condition at infinity. As in the unitary case we integrate over the Grassmann variables employing Theorem \ref{t1}. Integration over the group ${\rm SO}(2)$ in the boson--boson block yields
\begin{equation}\label{5.8}
 I(x_{11},x_{21},x_2)=\pi\int\limits_{\mathbb{R}}\int\limits_{\mathbb{R}}\int\limits_{\mathbb{R}} \phi_2^{(1)}(s_{11},s_{21},x_{11},x_{21})e^{-2e^{\imath\psi}s_2x_2}D_{s}^{(1,4;2,1)}f(s) ds_2ds_{11}ds_{21} ,
\end{equation}
where $\phi_2^{(1)}$ is the matrix Bessel function for ${\rm SO}(2)$. This matrix Bessel function can be expressed with the standard Bessel function,\cite{AbrSte72} see Ref. \onlinecite{GuhKoh02},
\begin{equation}\label{5.9}
\phi_2^{(1)}(s_{11},s_{21},x_{11},x_{21})= e^{(s_{11}+s_{21})(x_{11}+x_{21})/2}J_0\left(-\imath\frac{(s_{11}-s_{21})(x_{11}-x_{21})}{2}\right)
\end{equation}
with the normalization $J_0(0)=1$. We integrate by parts twice and define $\Phi(s,x)=$ $\phi_2^{(1)}(s_{11},s_{21},x_{11},x_{21})e^{-2 e^{\imath\psi}s_2x_2}$. We obtain
\begin{equation}\label{5.10}
 \begin{split}
   & I(x_{11},x_{21},x_2)=\\
   & = \frac{e^{2\imath\psi}}{2\pi}\int\limits_{\mathbb{R}}\int\limits_{\mathbb{R}}\int\limits_{\mathbb{R}} \frac{|s_{11}-s_{21}|}{s_{11}-s_{21}}\Phi(s,x)\times\\
   & \times\left(2\frac{\partial}{\partial s_{11}}+e^{-\imath\psi}\frac{\partial}{\partial s_2}\right)\frac{1}{s_{21}-e^{\imath\psi}s_2}\left(2\frac{\partial}{\partial s_{21}}+e^{-\imath\psi}\frac{\partial}{\partial s_2}\right)f(s)ds_2ds_{11}ds_{21}=\\
   & = 2\imath e^{\imath\psi}f(0)+\\
   & +2\imath e^{\imath\psi}\int\limits_{\mathbb{R}^+}\left.\left[f\left(\left(1-e^{-2\imath\psi}\right)\frac{\partial}{\partial z^*}+\frac{\partial}{\partial r}\right)\Phi-\Phi\left(\left(1-e^{-2\imath\psi}\right)\frac{\partial}{\partial z^*}-\frac{\partial}{\partial r}\right)f\right]\right|_{z=z^*=r}dr+\\
   & + \frac{e^{2\imath\psi}}{2\pi}\int\limits_{\mathbb{R}}\int\limits_{\mathbb{R}}\int\limits_{\mathbb{R}} \frac{|s_{11}-s_{21}|}{s_{11}-s_{21}}f(s)\times\\
   &\times \left(2\frac{\partial}{\partial s_{21}}+e^{-\imath\psi}\frac{\partial}{\partial s_2}\right)\frac{1}{s_{21}-e^{\imath\psi}s_2}\left(2\frac{\partial}{\partial s_{11}}+e^{-\imath\psi}\frac{\partial}{\partial s_2}\right)\Phi(s,x)ds_2ds_{11}ds_{21} .
 \end{split}
\end{equation}
We have used the permutation symmetry of both bosonic eigenvalues and the coordinate transformation of the proof of Theorem \ref{t6}. We are interested in the third summand of \eqref{5.10} because the integrand is the invariant function times the Berezinian $B_{2,2}^{(1,4)}(s_1,e^{\imath\psi}s_2)=|s_{11}-s_{21}|/\left((s_{11}-e^{\imath\psi}s_2)^2(s_{21}-e^{\imath\psi}s_2)^2\right)$ and a function. This function is the supermatrix Bessel function regarding to the group, $\UOSp(2/2)$\cite{GuhKoh02b}
\begin{equation}\label{5.11}
 \begin{split}
  &\Phi_{2,1}^{(1,4)}(s,x)= \\
  = & \frac{1}{2\pi}\frac{(s_{11}-e^{\imath\psi}s_2)^2(s_{21}-e^{\imath\psi}s_2)^2}{s_{11}-s_{21}}\left[\left(2\frac{\partial}{\partial s_{21}}+e^{-\imath\psi}\frac{\partial}{\partial s_2}\right)\frac{1}{s_{21}-e^{\imath\psi}s_2}\left(2\frac{\partial}{\partial s_{11}}+e^{-\imath\psi}\frac{\partial}{\partial s_2}\right)-\right.\\
  - & \left.\left(2\frac{\partial}{\partial s_{11}}+e^{-\imath\psi}\frac{\partial}{\partial s_2}\right)\frac{1}{s_{11}-e^{\imath\psi}s_2}\left(2\frac{\partial}{\partial s_{21}}+e^{-\imath\psi}\frac{\partial}{\partial s_2}\right)\right]\Phi(s,x)=\\
  = & \frac{1}{2\pi}\left[\left[\left(R-e^{\imath\psi}s_2\right)^2-r^2\right]\left[\left(\frac{\partial}{\partial R}+e^{-\imath\psi}\frac{\partial}{\partial s_2}\right)^2-\frac{\partial^2}{\partial r^2}\right]-2\left[R-e^{\imath\psi}s_2\right]\left[\frac{\partial}{\partial R}+e^{-\imath\psi}\frac{\partial}{\partial s_2}\right]-\right.\\
  - & \left.\frac{(R-e^{\imath\psi}s_2)^2+r^2}{r}\frac{\partial}{\partial r}\right]\Phi(R+r,R-r,s_2,x),
 \end{split}
\end{equation}
where we have changed the coordinates to $R=\frac{1}{2}(s_{11}+s_{21})$ and $r=\frac{1}{2}(s_{11}-s_{21})$. Now, we use the explicit representation \eqref{5.9} of the matrix Bessel function and the differential equation for the standard Bessel function $J_0$,
\begin{equation}\label{5.12}
 \left(\frac{\partial^2}{\partial r^2}+\frac{1}{r}\frac{\partial}{\partial r}+k^2\right)J_0(kr)=0\ .
\end{equation}
Thus, we find for \eqref{5.11}
\begin{equation}\label{5.13}
 \begin{split}
  \Phi_{2,1}^{(1,4)}(s,x)= & \frac{1}{2\pi}e^{R(x_{11}+x_{21})-2 e^{\imath\psi}s_2x_2}\left[\left[\left(R-e^{\imath\psi}s_2\right)^2-r^2\right]\left[\Str^2x+\frac{1}{r}\frac{\partial}{\partial r}-(x_{11}-x_{21})^2\right]-\right.\\
  -& \left.2\left[R-e^{\imath\psi}s_2\right]\Str x-\frac{(R-e^{\imath\psi}s_2)^2+r^2}{r}\frac{\partial}{\partial r}\right]J_0(-\imath r(x_{11}-x_{21}))=\\
  = & \frac{1}{2\pi}e^{R(x_{11}+x_{21})-2 e^{\imath\psi}s_2x_2}\left[4\left(s_{11}-e^{\imath\psi}s_2\right)\left(s_{21}-e^{\imath\psi}s_2\right)\left(x_{11}-x_2\right)\left(x_{21}-x_2\right)-\right.\\
  -& \left.\Str s\ \Str x-2r\frac{\partial}{\partial r}\right]J_0(-\imath r(x_{11}-x_{21})) .
 \end{split}
\end{equation}
Indeed, this is up to a constant the same result for the supermatrix Bessel function for $\UOSp(2/2)$ as in Ref. \onlinecite{GuhKoh02b}.  We remark that for the function
\begin{equation}\label{5.15}
 f(\sigma)= e^{-\Str\sigma^2}\Sdet^{-N/2}(\sigma+\imath\epsilon\e_4),
\end{equation}
its partial derivative
\begin{equation}\label{5.16}
 \left(2\frac{\partial}{\partial s_{j1}}+e^{-\imath\psi}\frac{\partial}{\partial s_{2}}\right)f(s)
\end{equation} 
vanishes at $s_{j1}=s_2=0$ for $j\in\{1,2\}$. Moreover, the first derivative of the standard Bessel function is zero at the point zero. Hence, we have the following integral representation of the Eq. \eqref{5.14}
\begin{equation}\label{5.17}
 \begin{split}
  Z(x+J)= & 1+C\frac{2e^{2\imath\psi}}{\pi}\int\limits_{\mathbb{R}}\int\limits_{\mathbb{R}}\int\limits_{\mathbb{R}} e^{-\Str(\sigma+x+\textbf{J})^2}\Sdet^{-N/2}(\sigma+\imath\epsilon\e_4)\times\\
  \times & \left(\frac{16J_1^2|s_{11}-s_{21}|}{(s_{11}-e^{\imath\psi}s_2)(s_{21}-e^{\imath\psi}s_2)}+\frac{4J_1\Str s|s_{11}-s_{21}|}{(s_{11}-e^{\imath\psi}s_2)^2(s_{21}-e^{\imath\psi}s_2)^2}\right)ds_2ds_{11}ds_{21} .
 \end{split} 
\end{equation}
This is indeed the correct generator, see Refs. \onlinecite{GuhKoh02b},\onlinecite{GGK04}, and \onlinecite{BreHik03}. We notice that without the properties of the invariant function, vanishing at zero and fulfilling \eqref{t6r}, we get two additional boundary terms, see \eqref{5.10}. These terms are needed to regularize the integral \eqref{5.10} at zero. If the invariant function is well behaved, for example $C^\infty$ in their supertraces, then these terms disappear.

\subsection{Supermatrix Bessel function for $\U(k_1/k_2)$}

We consider the supermatrix Bessel functions (\ref{b1}) with a generalized Wick rotation $e^{\imath\psi}$. The definition (\ref{b1}) is equivalent to the implicit definition
\begin{equation}\label{b2}
\begin{split}
 & \int\limits_{\Herm(2,k_1)}\int\limits_{\Herm(2,k_2)} \int f(\sigma)e^{\Str\sigma x}d[e^{-\imath\psi/2}\eta]d[e^{\imath\psi}\sigma_2]d[\sigma_1]=\\
 = & \int\limits_{\mathbb{R}^{k_1}}\int\limits_{\mathbb{R}^{k_2}}f(s)\Phi_{k_1k_2}^{(2,2)}(s,x)B_{k_1k_2}^{(2,2)}(s_1,e^{\imath\psi} s_2)d[e^{\imath\psi}s_2]d[s_1]+{\rm b.\ t.}
\end{split}
\end{equation}
for all rotational invariant functions $f$ with zero boundary condition at infinity. Up to boundary terms (b. t.) in the manifold of $\Herm(2,k_1)\dot{\oplus}\Herm(2,k_2)$ the integral on the left hand side of Eq. \eqref{b2} is equal with the integral on the right hand side. The exponential term on the left hand side does not depend on Grassmann variables. Thus, we shift the integral over these variables and use Theorem \ref{t1} and the operator $D_s^{(2,2;k_1k_2)}$ in Eq. \eqref{4.2}. We obtain
\begin{equation}\label{b3}
\begin{split}
 & e^{\imath\psi k_1k_2}\int\limits_{\Herm(2,k_1)}\int\limits_{\Herm(2,k_2)} e^{\Str\sigma x}D_s^{(2,2;k_1k_2)}f(s)d[e^{\imath\psi}\sigma_2]d[\sigma_1]=\\
 = & \int\limits_{\mathbb{R}^{k_1}}\int\limits_{\mathbb{R}^{k_2}}f(s)\Phi_{k_1k_2}^{(2,2)}(s,x)B_{k_1k_2}^{(2,2)}(s_1,e^{\imath\psi} s_2)d[e^{\imath\psi}s_2]d[s_1]+{\rm b.\ t.}
\end{split}
\end{equation}
We use the Itzykson--Zuber integral\cite{Har58,ItzZub80} for the boson--boson and fermion--fermion block and rewrite $D_s^{(2,2;k_1k_2)}$ with Eq. \eqref{4.7}. Then, we have
\begin{equation}\label{b4}
 \begin{split}
  & \frac{1}{\Delta_{k_1}(x_1)\Delta_{k_2}(x_2)}\int\limits_{\mathbb{R}^{k_1}}\int\limits_{\mathbb{R}^{k_2}} \det\left[e^{s_{m1} x_{n1}}\right]_{1\leq m,n\leq k_1}\det\left[e^{e^{\imath\psi}s_{m2} x_{n2}}\right]_{1\leq m,n\leq k_2}\sum\limits_{n=0}^{k_1k_2}\binom{k_1k_2}{n}\times\\
  &\times\left(\Str\frac{\partial^2}{\partial s^2}\right)^{k_1k_2-n}V_{k_1k_2}(s_1,e^{\imath\psi}s_2)\left(-\Str\frac{\partial^2}{\partial s^2}\right)^n\left(\sqrt{B_{k_1k_2}^{(2,2)}(s_1,e^{\imath\psi} s_2)}f(s)\right)d[s_2]d[s_1]=\\
  & =\frac{2^{k_1k_2}(k_1k_2)!\pi^{(k_1+k_2)/2}}{\pi^{(k_1-k_2)^2/2}}\int\limits_{\mathbb{R}^{k_1}}\int\limits_{\mathbb{R}^{k_2}}f(s)\Phi_{k_1k_2}^{(2,2)}(s,x)B_{k_1k_2}^{(2,2)}(s_1,e^{\imath\psi} s_2)d[s_2]d[s_1]+{\rm b.\ t.}
 \end{split}
\end{equation}
Due to the symmetry in the $x$ variables, we omit the determinants and get a factor of $k_1!k_2!$. Then, we partially integrate and act the differential operators onto the exponential functions. The emerging boundary terms are identified with these on the right hand side. Thus, we get
\begin{equation}\label{b5}
 \begin{split}
  & \frac{1}{\Delta_{k_1}(x_1)\Delta_{k_2}(x_2)}\int\limits_{\mathbb{R}^{k_1}}\int\limits_{\mathbb{R}^{k_2}}\sqrt{B_{k_1k_2}^{(2,2)}(s_1,e^{\imath\psi} s_2)}f(s)\times\\
  &\times\sum\limits_{n=0}^{k_1k_2}\binom{k_1k_2}{n}\left(\Str\frac{\partial^2}{\partial s^2}\right)^{k_1k_2-n}V_{k_1k_2}(s_1,e^{\imath\psi}s_2)\left(-\Str\frac{\partial^2}{\partial s^2}\right)^ne^{\Str sx}d[s_2]d[s_1]=\\
  = & \frac{1}{\Delta_{k_1}(x_1)\Delta_{k_2}(x_2)}\int\limits_{\mathbb{R}^{k_1}}\int\limits_{\mathbb{R}^{k_2}}\sqrt{B_{k_1k_2}^{(2,2)}(s_1,e^{\imath\psi} s_2)}f(s)e^{\Str sx}\times\\
  & \times\left(\Str\frac{\partial^2}{\partial s^2}+2\sum\limits_{n=1}^{k_1}x_{n1}\frac{\partial}{\partial s_{n1}}+2e^{-\imath\psi}\sum\limits_{n=1}^{k_2}x_{n2}\frac{\partial}{\partial s_{n2}}\right)^{k_1k_2}V_{k_1k_2}(s_1,e^{\imath\psi}s_2)d[s_2]d[s_1]=\\
  = & \frac{2^{k_1k_2}(k_1k_2)!\pi^{(k_1+k_2)/2}}{k_1!k_2!\pi^{(k_1-k_2)^2/2}}\int\limits_{\mathbb{R}^{k_1}}\int\limits_{\mathbb{R}^{k_2}}f(s)\Phi_{k_1k_2}^{(2,2)}(s,x)B_{k_1k_2}^{(2,2)}(s_1,e^{\imath\psi} s_2)d[s_2]d[s_1] .
 \end{split}
\end{equation}
The differential operator acts on a polynomial of order $k_1k_2$. An expansion of this operator leads to a differential operator which depends on a sum of derivatives of order $k_1k_2$ to $2k_1k_2$. Therefore, the second derivatives do not contribute. The remaining operator acts on the polynomial and we find
\begin{equation}\label{b6}
\begin{split}
 & \int\limits_{\mathbb{R}^{k_1}}\int\limits_{\mathbb{R}^{k_2}}\sqrt{\frac{B_{k_1k_2}^{(2,2)}(s_1, e^{\imath\psi} s_2)}{B_{k_1k_2}^{(2,2)}(x_1,x_2)}}f(s)e^{\Str sx}d[s_2]d[s_1]=\\
  = & \frac{2^{k_1k_2}\pi^{(k_1+k_2)/2}}{k_1!k_2!\pi^{(k_1-k_2)^2/2}}\int\limits_{\mathbb{R}^{k_1}}\int\limits_{\mathbb{R}^{k_2}}f(s)\Phi_{k_1k_2}^{(2,2)}(s,x)B_{k_1k_2}^{(2,2)}(s_1,e^{\imath\psi} s_2)d[s_2]d[s_1] .
\end{split}
\end{equation}
Now, we analyze both integrals for all rotational invariant functions $f$. Thereby, we take notice of the invariance of $f$ regarding to the tensor product of the permutation group $S(k_1)\otimes S(k_2)$ acting on the boson--boson and the fermion--fermion block. We get 
\begin{equation}\label{b7}
 \Phi_{k_1k_2}^{(2,2)}(s,x)=\frac{\pi^{(k_1-k_2)^2/2}}{2^{k_1k_2}\pi^{(k_1+k_2)/2}}\frac{\det\left[e^{s_{m1} x_{n1}}\right]_{1\leq m,n\leq k_1}\det\left[e^{e^{\imath\psi}s_{m2} x_{n2}}\right]_{1\leq m,n\leq k_2}}{\sqrt{B_{k_1k_2}^{(2,2)}(s_1,e^{\imath\psi} s_2)B_{k_1k_2}^{(2,2)}(x_1, x_2)}}.
\end{equation}
Indeed, this is for $k_1=k_2=k$, $e^{\imath\psi}=\imath$, and exchanging $s\to\imath s$ the correct result.\cite{Guh96,Guh96b,Guh06} Also, we notice that the choice of the normalization constant in the measure $d\mu(U)$ arises in a natural way if we take the Eq. \eqref{b2} as the definition of the supermatrix Bessel functions.\cite{Guh96b,Guh06}

\section{Remarks and conclusions}

We derived a handy form for the differential operator with respect to commuting variables acting on an invariant superfunction. This operator is equivalent to integrating Grassmann variables over the same function. It is uniquely defined by the invariance class which the function fulfills. Detailed group theoretical considerations are not needed in our approach. We only used a mapping from the whole superspace to $(\Lambda^0(p,2L))^p$ which leaves the superfunction invariant. Here, an important remark is in order. Various invariant superfunctions exist which depend on a number of independent invariants larger than the dimension of the superspace body. The integration over the Grassmannians gives a differential operator with respect to these invariants as independent variables. Otherwise, a differential operator regarding to the commuting elements in $(\Lambda^0(p,2L))^p$ could not exist. We expect that there is a superior supermanifold of the original one, $L(p,2L)$, for which Theorem \ref{t1} would be valid as well. As an example, we consider a $\U(k_1/k_2)$--invariant superfunction which depends on a complex supervector $v$ and a $\U(k_1/k_2)$--symmetric supermatrix $\sigma$. The independent invariants are the length of the supervector, the supertraces from the first to the $(k_1+k_2)$th power of the supermatrix, and the expectation values $v^\dagger\sigma^j v$, $j\in\{1,k_1+k_2-1\}$. The resulting differential operator must be related to the operators which we have obtained in the Sec. III and IV for the group $\U(k_1/k_2)$. An extension of Theorem \ref{t1} would be a helpful tool to calculate the $k$--point correlation function of rotational invariant ensembles. One does not know what the ordinary matrix Bessel functions explicitly look like. However, one can carry out the calculation to integrals over invariant superfunctions with a large number of invariants.

There are various strong motivations for deriving formula \eqref{t1.1}. First of all, we aim at giving an explicit transformation formula in contrast to Rothstein\cite{Rot87} of the change from Cartesian coordinates of a matrix to the eigenvalue--angle coordinates. The full account of the Efetov--Wegner terms is closely linked with this task and is also an aim of this work. As shown in many studies,\cite{Guh06,Zir91,MirFyo91,DFZ92} these terms guarantee the normalization or rather the reduction to a smaller integral if the function is whole or partly invariant under the action of a supergroup. As obvious from Refs. \onlinecite{BasAke07} and \onlinecite{DFZ92,MirFyo91,Zir91,EST04,EfeKog04,LSZ07,Som07,BEKYZ07}, an extension of this theory to curved Riemannian superspaces is also desirable. Such a theorem would be very useful for these cases to study also the superbosonization formula\cite{EST04,EfeKog04,LSZ07,Som07,BEKYZ07,BasAke07} in random matrix theory.

\section*{Acknowledgement}
We acknowledge support from Deutsche Forschungsgemeinschaft within Sonderforschungsbereich Transregio 12.
H.K. acknowledges support from Deutsche Forschungsgemeinschaft with Grant No. 3538/1-1.
\appendix

\section{On certain integrals for functions invariant under UOSp(1/2)}

The differential operator which results from an integration over the Grassmann variables in the matrix case of the supergroup $\UOSp(1/2)$ for the representation $\UOSp^{(+)}(1/2)$ is given by \eqref{4.24}. For the other representation we get a similar expression in which the factor of 2 stands in front of the fermionic partial derivative. We consider the integral
\begin{equation}\label{a1.1}
 I[f,\alpha]=\int\limits_{\mathbb{R}^2}\frac{1}{x-e^{\imath\psi}y}\left(\frac{\partial}{\partial x}+\alpha e^{-\imath\psi}\frac{\partial}{\partial y}\right)f(x,y)dxdy
\end{equation}
for ${\rm Re}\left(e^{2\imath\psi}/\alpha\right)<0$, where $f$ is a Schwartz function, analytic, and $I[f,\alpha]$ finite for all $\alpha$ which fulfills the first requirement. $I$ has the properties
\begin{eqnarray}
 I\left[{\rm exp}\left(-x^2+\frac{e^{2\imath\psi}}{\alpha}y^2\right),\alpha\right] & = & -2\pi\imath e^{-\imath\psi}\sqrt{\alpha}=C_\alpha,\label{a1.2}\\
 I[f,\alpha]+I[f,\beta] & = & 2I\left[f,\frac{\alpha+\beta}{2}\right],\label{a1.3}\\
 I[f,1] & = & -2\pi\imath e^{-\imath\psi} f(0).\label{a1.4}
\end{eqnarray}
The first and the third property are obvious. Due to the linearity of the integral, the second one is true if $I[f,\alpha]$ and $I[f,\beta]$ exist. Surprisingly, there is no Cauchy--like integral theorem for $\alpha\neq1$. The integral $I[f,1]$ represents the $\U(1/1)$--case. More precisely, we have
\begin{theorem}\label{t8}\ \\
 There is no $\alpha\neq1$ with ${\rm Re}\left(e^{2\imath\psi}/\alpha\right)<0$ such that $I[f,\alpha]={\rm const.}f(0)$ for all Schwartz functions which are analytic and possess a finite $I[f,\alpha]$.
\end{theorem}
\textbf{Proof:}\\
We assume that there exists an $\alpha\neq 1$ with the described requirements which fulfill $I[f,\alpha]={\rm const.} f(0)$. Then, the constant is equal to $C_\alpha$ because the Gaussian function in \eqref{a1.2} fulfills the requirements of the function in the theorem. Thus, we use \eqref{a1.2} and \eqref{a1.4}. We compute
\begin{equation}\label{a1.5}
 I\left[f,\frac{\alpha+1}{2}\right]\overset{\eqref{a1.3}}{=}\frac{1}{2}(I[f,\alpha]+I[f,1])=\frac{1}{2}(C_\alpha+C_1)f(0) .
\end{equation}
Therefore, there exists a constant for $I\left[f,(\alpha+1)/2\right]$ and this constant is equal to $(C_\alpha+C_1)/2$. On the other hand, the constant is unique and, accordingly, equal to $C_{(\alpha+1)/2}$. We find
\begin{equation}\label{a1.6}
 C_{(\alpha+1)/2}=\frac{1}{2}(C_\alpha+C_1),
\end{equation}
which becomes after some calculation
\begin{equation}\label{a1.7}
 (\alpha-1)^2=0 .
\end{equation}
As this contradicts the assumption, the theorem is proven.\hfill$\square$

We give a counterexample to illustrate this theorem. We consider the function
\begin{equation}\label{a1.8}
 f(x,y)=\left(x-\frac{e^{\imath\psi}}{\alpha}y\right)^2e^{-x^2+e^{2\imath\psi}y^2/\alpha} ,
\end{equation}
which vanishes at zero. However, the integral is
\begin{equation}\label{a1.9}
 I[f,\alpha]=-\imath\pi e^{-\imath\psi}\sqrt{\alpha}\left(1-\frac{1}{\alpha}\right)\overset{\alpha\neq1}{\neq}0 .
\end{equation}

Consequently Efetov's method\cite{Efe83} to derive such integral theorems cannot be applied to all kinds of integrals over invariant functions on superspaces. Nevertheless, it is a mystery to us why this method works for $\U(k/k)$ and $\UOSp(2k/2k)$--symmetric matrices, see Theorems \ref{t5}, \ref{t6} and \ref{t7}, but would fail for $\UOSp(2k-1/2k)$, even though there is the same number of anticommuting and commuting variables to integrate.

\end{document}